\documentclass[onecolumn]{aastex631}
\usepackage{bm}

\usepackage[outdir=./]{epstopdf}

\shorttitle{Impact-induced atmosphere loss}
\shortauthors{Kurosaki \& Inutsuka}

\graphicspath{{./}{figures/}}

\begin{document}

\title{Giant Impact Events for Protoplanets: Energetics of Atmospheric Erosion by Head-on Collision}

\author[0000-0002-9958-176X]{Kenji Kurosaki}
\affiliation{Department of Physics, Graduate School of Science, Nagoya University, Furo-rho, Chikusa-ku, Nagoya, Aichi, 464-8601, Japan}
\affiliation{Graduate School of Science, Kobe University, 1-1, Rokkkodai-cho, Nada-ku, Kobe, Hyogo 657-8501, Japan}
\email{kenji.kurosaki@people.kobe-u.ac.jp}
\correspondingauthor{Kenji Kurosaki}

\author{Shu-ichiro Inutsuka}
\affiliation{Department of Physics, Graduate School of Science, Nagoya University, Furo-rho, Chikusa-ku, Nagoya, Aichi, 464-8601, Japan}

\begin{abstract}
Numerous exoplanets with masses ranging from Earth to Neptune and radii larger than Earth have been found through observations. These planets possess atmospheres that range in mass fractions from 1\% to 30\%, reflecting the diversity of atmospheric mass fractions.
Such diversities are supposed to be caused by differences in the formation processes or evolution. Here we consider 
head-on giant impacts onto planets
causing atmosphere losses in the later stage of their formation.
We perform smoothed particle hydrodynamic simulations to study the impact-induced atmosphere loss of young super-Earths with 10\%--30\% initial atmospheric mass fractions. 
We find that the kinetic energy of the escaping atmosphere is almost proportional to the sum of the kinetic impact energy and self-gravitational energy released from the merged core. 
We derive the relationship between the kinetic impact energy and the escaping atmosphere mass.
The giant impact events for planets of comparable masses are required in the final stage of the popular scenario of rocky planet formation.
We show it results in a significant loss of the atmosphere,  if the impact is a head-on collision with comparable masses.
This latter fact provides a constraint on the formation scenario of rocky planets with substantial atmospheres.
\end{abstract}

\keywords{Super Earths (1655), Exoplanet atmospheres (487), Impact phenomena (779), Hydrodynamical simulations (767)}

\section{Introduction} \label{sec:intro}
The standard scenario of terrestrial planets' formation invokes the formation of planetesimals in the protoplanetary disk formed along with a central protostar. The protoplanets, resulting in the aggregation of planetesimals embedded in the protoplanetary disk, acquire a gaseous envelope called the primordial atmosphere \citep[e.g.,][]{Ikoma2006,Ormel2015}.
As a result, protoplanets with primordial atmospheres are formed in the protoplanetary disk. When planets collide, their atmospheres should receive impact energies,
and certain amounts of atmospheres tend to escape
through a giant impact \citep{Stewart2014, Shuvalov2014,Liu2015b,Ginzburg2018,Schlichting2018,Yalinewich2019, Biersteker2019,Denman2020,Denman2022,Kegerreis2020a,Kegerreis2020b}.
Furthermore, a giant impact can change the composition of the primordial atmosphere by chemical reactions led by the high-temperature environments resulting from the collision \citep[e.g.,][]{Massol2016}.
That event has a potential to alter the planetary surface environment \citep{Abe1986,Kurosawa2015,Sakuraba2021}.
The atmospheric properties are important in characterizing the properties of an exoplanet. Recent studies suggest that a planetary radius larger than $1.6R_\oplus$ is expected to have an atmosphere \citep[e.g.,][]{Rogers2015,Fulton2018}.
In this study, we focus on super-Earths with masses and radii smaller than 10 times that of the Earth. Those planets have typically 10\%--30\% of atmospheres by mass.
Although recent research has indicated that exoplanets might acquire atmospheres during the formation phase, there are still unknowns regarding the disk lifetime and atmospheric opacity, making it difficult to predict the mass of the ensuing atmosphere immediately after the formation \citep{Ikoma2012,Venturini2017}.
After the formation, those planets might undergo impact events causing an atmosphere loss \citep{Poon2020,Ogihara2021,Matsumoto2021}.
The atmosphere loss induced by a giant impact has been studied to investigate the origin of the primordial atmosphere of the early Earth \citep{Genda2003,Shuvalov2014,Inamdar2015}, other terrestrial planets or satellites \citep{Korycansky2011}, and the giant planet \citep{Liu2015b}.
Furthermore, the giant impact can reset the initial atmosphere impacting the atmospheric accretion history \citep{Ali-Dib2021arXiv}, or
produce impact-induced debris \citep{Genda2015,Schneiderman2021}.
That is, the giant impact event is one of the important phenomena that affect the character of the planet.

In this study, we focus on the impact-induced atmosphere loss of early age super-Earths that have massive atmospheres.
\citet{Inamdar2015} and \citet{Biersteker2019} studied
the atmospheric expansion after the giant impact where the ground motion induces the shock wave and the atmospheric loss induced by the shock wave.
It is necessary to consider both atmospheric heating and ground motion.  \citep[e.g.,][]{Ahrens1993,Genda2003,Schlichting2015}.
\citet{Kegerreis2020a} and \citet{Denman2020} have calculated 3-dimensional hydrodynamic simulation to investigate the impact-induced atmosphere loss.

\citet{Kegerreis2020a} calculated the direct interaction of the large impactor and analyzed the instant loss mechanism.
They showed that the acceleration of the ground motion causes the atmospheric outflow's velocity to the escape velocity of the planet immediately after the impact.
This corresponds to the impact-induced motion of the core driving the atmospheric loss.
\citet{Kegerreis2020b} calculated giant impact simulations for several Earth-mass planets whose atmospheric mass fraction is 1\%. They calculated head-on and oblique impacts and found a scaling law to predict the atmosphere loss.
\citet{Denman2020} performed the head-on giant impact simulation of super-Earths with 10\%--30\% atmospheric masses. They showed that the atmosphere loss fraction of a planet with 30\% of an atmosphere is less than that of a planet with 10\% of an atmosphere when those planets obtain the same specific relative kinetic energy of the impact. They also found that the critical impact energy changed the scaling law for the atmospheric escape.
\citet{Denman2020,Denman2022} used the specific kinetic energy $Q_r=\mu v_\mathrm{imp}^2/(2M_\mathrm{tot}) $ to estimate atmospheric loss.
On the other hand, \citet{Kegerreis2020a,Kegerreis2020b} considered that the atmospheric loss is proportional to $(v_\mathrm{imp}/v_\mathrm{esc})^2 (M_i/M_\mathrm{tot})^{1/2} (\rho_I/\rho_T)^{1/2}$. 
Since the scaling laws for thin and thick initial atmospheres differ, we should not expect their results to be identical scaling law relations.
Studies by \citet{Kegerreis2020b} and \citet{Denman2022} suggested that the impact-induced atmosphere loss is dominated by the impact energy, which may be the same as in the 1-D calculation.
It remains to investigate the apparent difference in their arguments.

In this article, we focus on the atmosphere loss from super-Earths.
Specifically, we consider the head-on collision for super-Earth with an atmosphere by a comparable mass of impactor with or without an atmosphere.
Three dimensional impact simulations \citep{Kegerreis2020b,Denman2022} have suggested energy scaling laws in impact events that may relate to the energy distribution of the atmosphere and core.
This study focuses only on head-on collisions and performs 3-dimensional numerical simulations. Furthermore, we analyze the energy distribution of the head-on collision simulation to predict the impact-induced atmosphere loss using the analytically derived energy distribution of the atmosphere and core.
We calculate giant impact simulations for a super-Earth with an atmosphere of 10\%--30\% by mass and discuss the energy scaling law for the impact-induced atmosphere loss based on the energy conservation law.
The energy distribution before and after the impact is determined by numerical simulation, and the scaling law of the impact-induced atmosphere loss is analyzed.
In section \ref{sec:method}, we describe the numerical method and settings.
In section \ref{sec:results}, we show examples of our numerical simulations.
In section \ref{sec:anal}, we show the analysis results of our numerical simulations.
In section \ref{sec:discussion}, we discuss the implication of the results for the formation history of the planet with a massive atmosphere.
Finally, in section \ref{sec:conclusion}, we conclude our study.

\section{Method} \label{sec:method}
In this section, we summarize the numerical method to study the head-on, impact-induced atmosphere loss.
In our simulation, we use the smoothed particle hydrodynamics simulation \citep[e.g.,][here after SPH]{Lucy1977,Monaghan1992} 
to solve the following hydrodynamic equations:
\begin{eqnarray}
\frac{d\rho}{dt} &=& -\rho \nabla\cdot\bm{v} \\
\frac{d\bm{v}}{dt} &=& -\frac{1}{\rho} \nabla P + \nabla \int dx'^3 \frac{G\rho(x')}{|\bm{x}-\bm{x'}|} \\
\frac{du}{dt} &=& -\frac{P}{\rho} \nabla \cdot\bm{v} \\
P &=& P(\rho, u)  \label{eos}
\end{eqnarray}
where $\rho, P, \bm{v}$ and $u$ are density, pressure, velocity, and specific internal energy, respectively.
$t$ is the time, $\bm{x}$ is the position, and
$G(=6.67408\times10^{-11}~\mathrm{m}^3~\mathrm{kg}^{-1}~\mathrm{s}^{-2})$
is the gravitational constant.
We normalize $t$ by the overall impact event using the target's free-fall time described as
\begin{equation}
\tau_\mathrm{ff} = \sqrt{\frac{3\pi}{32 G \rho_T}}, \label{tauff-def}
\end{equation}
where $\rho_T$ is the mean density of the target.
We calculate the impact simulation until the evolution time reaches $t=50~\tau_\mathrm{ff}$.
For example, we choose a target with mass of 1 Earth-mass with 10\% of an atmosphere by mass, while the impactor is the same (see Table~\ref{planetary_model}).
In this case, we stop the calculation when the evolution time reaches $t=110$ hours since $\tau_\mathrm{ff}$ is 2.2~hours.

We have implemented the acceleration modules for our SPH code with FDPS \citep{Iwasawa2016} and FDPS Fortran interface \citep{Namekata2018}. 
We adopt the SPH code developed by \citet{Kurosaki2019}.
Our SPH code is based on the standard smoothed particle hydrodynamics method. We adopt the Von Neumann-Richtmyer viscosity for artificial viscosity \citep[e.g.,][]{Monaghan1992}.

The planetary structure is composed of a rocky core surrounded by a hydrogen-helium atmosphere.
We use the basalt \citep{Pierazzo2005, Collins2016} equation of state constructed using M-ANEOS \citep{Thompson1972,Melosh2007,Thompson2019} for a rock core.
Furthermore, we use \citet{Saumon1995} for a hydrogen-helium atmosphere.
Table~\ref{planetary_model} shows the planetary mass, the atmosphere mass fraction, and the radius relationship.
We set the mass of one SPH particle $10^{-5}~M_\oplus$.
The hydrogen particle mass is the same as the rock particle mass.
In the case of $1~M_\oplus$ target with 30\% of the atmosphere, we use $3 \times 10^{4}$ particles for $0.3~M_\oplus$ hydrogen atmosphere and $7 \times 10^{4}$ particles for $0.7~M_\oplus$ rocky core.

The detailed model of the planetary structure is shown in \citet{Kurosaki2017}.
Here, we summarize the method to determine the planetary structure. We calculate a spherically symmetric hydrostatic equilibrium structure. The planetary temperature structure is an adiabatic profile. We set the planetary entropy $S=S(P_\mathrm{surf}, T_\mathrm{surf})$, where $P_\mathrm{surf}$ and $T_\mathrm{surf}$ are the surface pressure and temperature, respectively. We set $P_\mathrm{surf}=100~$bar.
We calculate the time evolution of the fully convective planets to determine the entropy at 10 Myrs.
The planetary entropy values are $S\sim 6\times10^{4}~$J$\cdot$K$^{-1}\cdot$kg$^{-1}$. $T_\mathrm{surf}$ values are also shown in Table~\ref{planetary_model}. We assume the surface temperature for the planet without an atmosphere is $T(P=100~\mathrm{bar})=1500~$K. After getting the planetary interior structure, we determine SPH particle configurations using the relaxation method.
Initially, SPH particles are placed on a spherical shell, and relaxation is performed so that the density is consistent with the internal structure
\citep[e.g.,][]{Reinhardt2017,Ruiz-Bonilla2021}.
Next, we input the internal energy profile to SPH particles and calculate the hydrodynamic equations.
In our study, we calculate $5 \tau_\mathrm{ff}$$(\approx$ 10~hours) to determine the target structure.
After the relaxation, the target's density and internal energy structure are stable and well behaved.
The relative error of the density structure created by SPH particles with respect to the spherically symmetric hydrostatic equilibrium structure is less than 0.1~\%.
Note that we set a two-layered target. Previous studies \citep[e.g.,][]{Kegerreis2020a,Kegerreis2020b, Denman2020, Denman2022} have considered a three-layered planet composed of an iron core, a rock mantle, and an atmosphere from the bottom to the top. Since we aim to study the energy distribution between the atmosphere and the core, we set a two-layered target composed of a rock core and an atmosphere for simplicity.
Our SPH code shows the error of the angular momentum conservation is smaller than 0.1\%. On the other hand, the error of the energy conservation is smaller than 5\%, mainly caused by the gravitational part. Without the gravitational calculation, the angular momentum and energy conservation errors are less than 0.1\%  \citep[see also sec 3.4,][]{Inutsuka2002}.
We check the atmosphere loss for $10^4$, $10^5$, and $10^6$ particles for $1~M_\oplus$ with 30\% of the atmosphere. Then we calculate the impact simulation for the same body impact
and find $X_\mathrm{atm}=0.69$ for $10^4$, $X_\mathrm{atm}=0.72$ for $10^5$, and $X_\mathrm{atm}=0.73$ for $10^6$.
 We find that the atmosphere loss fraction converges 2\%. We use the $10^5$ particles for all other runs.

We focus on the total thermal energy and gravitational energy described by
\begin{eqnarray}
U &=& \int_{0}^{M} u~dM_r \label{U-def}\\
\Phi &=&-\int_{0}^{M} \frac{GM_r}{r}dM_r \label{Phi-def}
\end{eqnarray}
where $U$ is the total thermal energy, $\Phi$ is the gravitational energy, $u$ is the internal energy per unit mass, $r$ is the distance from the planetary center, and $M_r$ is the mass coordinate at $r$.
We use equal mass particles in our SPH simulations.
We simply define the thermal energy and gravitational energy of the atmosphere and the core shown by
\begin{eqnarray}
U_\mathrm{c} &=& \int_{0}^{M_c}  u~dM_r \label{Uc-def} \\
U_\mathrm{atm} &=& \int_{M_c}^{M} u~dM_r \label{Uatm-def} \\
\Phi_\mathrm{c} &=& -\int_{0}^{M_c}  \frac{GM_r}{r}dM_r \label{Phic-def} \\
\Phi_\mathrm{atm} &=& \Phi - \Phi_\mathrm{c}, \label{Phiatm-def}
\end{eqnarray}
where $M$ is the planetary mass, $M_c$ is the planetary core mass.
In this study, we consider a head-on collision to study the energy distribution between the kinetic energy of escaping mass and the heating energy of the gravitationally bound body after the impact.
The impact velocity is 1.0--2.3 $v_\mathrm{esc}$ where $v_\mathrm{esc}$ is an escape velocity described as
\begin{equation}
v_\mathrm{esc}=\sqrt{ \frac{2G(M_\mathrm{T}+M_\mathrm{I})}{R_\mathrm{T}+R_\mathrm{I}} }  \label{vesc-def}
\end{equation}
where $M_\mathrm{T}$ and $R_\mathrm{T}$ are the target's mass and radius, $M_\mathrm{I}$ and $R_\mathrm{I}$ are the impactor's mass and radius, respectively.
We define the kinetic impact energy as
\begin{equation}
K_\mathrm{imp} = \frac{\mu}{2} v_\mathrm{rel}^2 - \frac{GM_\mathrm{T} M_\mathrm{I}}{L}  \label{Kimp-def}.
\end{equation}
where $\mu (= M_\mathrm{T} M_\mathrm{I}/(M_\mathrm{T}+M_\mathrm{I}))$ is the reduced mass, $v_\mathrm{rel}$ is the relative velocity at a distance, $L$ where $L$ is the initial separation between the target and the impactor. 
Note that $K_\mathrm{imp}$ includes the gravitational energy induced by the target and the impactor interaction.
During the SPH simulation, we check particles' kinetic and gravitational energies for each time step.
We define escaping particles as those with
\begin{equation}
 \frac{m_i v_i^2}{2} - \sum_{j=1}^{N} \frac{Gm_j m_i}{|\mathrm{r}_i-\mathrm{r}_j|} \ge 0, \label{ejectcond}
\end{equation}
where $m_i$ and $m_j$ are the $i$-th and $j$-th particle masses, $r_i$ and $r_j$ are the positions of $i$-th and $j$-th particles, and $v_i$ is the velocity of the $i$-th particle, respectively.
The total energy of escaping particles is calculated by 
\begin{equation}
K_\mathrm{esc} + U_\mathrm{esc} =\sum_{i} \frac{1}{2} m_iv_i^2 + \sum_{i} m_i u_i \label{Kesc-def}
\end{equation}
where $K_\mathrm{esc}$ and $U_\mathrm{esc}$ are the total kinetic and internal energies of escaping particles, and $i-$th particles are only escaping particles.
We consider the escaping mass fraction for the atmosphere and the core described by
\begin{eqnarray}
X_\mathrm{atm} &=& \frac{M_\mathrm{esc,atm}}{M_\mathrm{T,atm}+M_\mathrm{I,atm}}, \label{Hloss} \\
X_\mathrm{core} &=& \frac{M_\mathrm{esc,c}}{M_\mathrm{T,c}+M_\mathrm{I,c}} \label{Rloss}
\end{eqnarray}
where $M_\mathrm{esc,atm}, M_\mathrm{esc,c}$ are escaping masses of the atmosphere and the core, $M_\mathrm{T,atm}, M_\mathrm{I,atm}$ are the atmospheric masses of the target and the impactor, and $M_\mathrm{T,c}, M_\mathrm{I,c}$ are the core masses of the target and the impactor. 
We describe subscript $\mathrm{B}$ as the gravitationally bound body.
The atmospheric mass fraction after the impact $Y_\mathrm{B,atm}$ is determined by the mass fraction of the gravitationally bound atmospheric mass, $M_\mathrm{B,atm}$, to the gravitationally bound planetary mass $M_\mathrm{B}$.
That is, we can describe $M_\mathrm{B}$ and $Y_\mathrm{B,atm}$ as
\begin{eqnarray}
M_\mathrm{B} &=& M_\mathrm{T}+M_\mathrm{I} - M_\mathrm{esc},  \label{Mb-def} \\  
Y_\mathrm{B} &=& \frac{M_\mathrm{atm}-M_\mathrm{esc,atm}}{M_\mathrm{B}}, \label{Yb-def}
\end{eqnarray}
where $M_\mathrm{atm}$ is the total atmospheric mass defined as $M_\mathrm{atm}=M_\mathrm{T,atm}+M_\mathrm{I,atm}$, and $M_\mathrm{esc}$ is the total escaping mass shown as $M_\mathrm{esc} =M_\mathrm{esc,atm} + M_\mathrm{esc,c}$.
Next, we describe the internal structure of the gravitationally bound body.

We stop an impact simulation when the atmosphere loss fraction becomes constant. 
This study adopts the numerical simulation results at  $t=50~\tau_\mathrm{ff}$ to analyze the post-impact internal structure.
When the time is $t=50~\tau_\mathrm{ff}$, we study the internal structure focusing on the gravitationally bound particles. 
We consider the center the highest density point of the gravitationally bound particles. 
We sort particles into 50 spherical shells from the inside to the outside of the gravitationally bound body in the whole simulation box.
Let $\rho_i$ and $u_i$ denote the density and the internal energy at the $i$-th spherical shell.
We define the number of particles in the i-th spherical shell at ($r_i, r_i+dr$] as $N_{r_i}$.
We calculate $\overline{\rho}_i$ and $\overline{u}_i$ as
\begin{eqnarray}
\overline{\rho}_i &=& \frac{\sum_{j=1}^{N_{r_j}} \rho_j ( r_i < |\mathbf{x}| \le r_i +\delta r_i )}{N_r} \\
\overline{u}_i &=& \frac{\sum_{j=1}^{N_{r_j}} u_j ( r_i < |\mathbf{x}| \le r_i +\delta r_i )}{N_{r_j}} 
\end{eqnarray}
where $\mathbf{x}$ is the position with origin at the center of the sphere, $r_i$ is the distance from the center of the $i$-th spherical shell, $\delta r_i$ is the thickness of the $i$-th spherical shell, and $\rho_j(\mathbf{x})$ and $u_j(\mathbf{x})$ are the density and internal energy of $j$-the particle at the position $\mathbf{x}$.
For the $i$-th spherical region, we evaluate the hydrogen mass fraction $y_B(r_i)$ calculated by
\begin{equation}
y_\mathrm{B}(r_i) = \frac{\sum_{k=1}^{N_{r,\mathrm{atm}}} m_{k,\mathrm{atm}} }{\delta M_r}, \label{Ybr-def}
\end{equation}
where $N_{r,\mathrm{atm}}$ is the number of hydrogen particles, $m_{k,\mathrm{atm}}$ is the mass of the $k$-th hydrogen particle, and $\delta M_r$ is the mass of the spherical shell where $\delta M_r = M_{r_i+\delta r_i} – M_{r_i}$.
In our calculation, Eq.~\ref{Ybr-def} is simply the number ratio of the atmosphere particles to the sum of the atmosphere and core particles.
The total internal and gravitational energy are calculated by Eq.~\ref{U-def} and Eq.~\ref{Phi-def}.
We divide the results into two categories: a layered structure and a mixed structure.
In this study, we assumed a planet with a mixed structure as the mixed ratio of the planet $y_\mathrm{B}(r=R_\mathrm{T}+R_\mathrm{I})$ is less than 0.9.
When we want to know whether the post-impact structure is mixed or not, we should analyze more strictly for the post-impact simulation because the interior structure of the post-impact planet will evolve. In this paper, we define the mixed mass fraction at $r=R_\mathrm{T}+R_\mathrm{I}$ and whether the post-impact planet is mixed well.

\startlongtable
\begin{deluxetable*}{cccccccccc}
\tablecaption{Properties of targets and impactors. We show the planetary mass $M~[M_\oplus]$, atmospheric mass fraction $Y$, temperature at 100 bar $T_\mathrm{surf}$, radius $R~[R_\oplus]$, core radius $R_c~[R_\oplus]$, internal energy of the core $U_c$ [$10^{38}$~erg], gravitational energy of the core $\Phi_c$ [$10^{38}$~erg], internal energy of the atmosphere $U_\mathrm{atm}$ [$10^{38}$~erg], and gravitational energy of the atmosphere $\Phi_\mathrm{atm}$ [$10^{38}$~erg].
\label{planetary_model}}
\tablewidth{700pt}
\tabletypesize{\scriptsize}

\tablehead{
\colhead{Name} & \colhead{$M~[M_\oplus]$} &  \colhead{$Y$} & \colhead{$T_\mathrm{surf}~[\mathrm{K}]$} & \colhead{$R~[R_\oplus]$} & \colhead{$R_c~[R_\oplus]$} & \colhead{$U_c$ [$10^{38}$~erg]} & \colhead{$-\Phi_c$ [$10^{38}$~erg]} & \colhead{$U_\mathrm{atm}$ [$10^{38}$~erg]} & \colhead{$-\Phi_\mathrm{atm}$ [$10^{38}$~erg]}
}
\startdata
A1 & 0.10           & 0.1  & 660 &  2.1\,\;  & 0.56      &    9.67$\times10^{-2}$ & 3.33$\times10^{-1}$  & 1.61$\times10^{-2}$ & 2.09$\times10^{-2}$ \\
A2 & 0.30           & 0.1 & 810  & 3.0\,\;  & 0.81      &    5.72$\times10^{-1}$   &  2.14\;\;\;\;\;\;\;\;\;\;\,  & 1.21$\times10^{-1}$  & 1.43$\times10^{-1}$  \\
A3 & 1.0\,\;       & 0.1 & 1080 & 4.3\,\;  &  1.2\,\;  &    4.46\;\;\;\;\;\;\;\;\;\;\,   &  1.64$\times10^{+1}$  & 8.36$\times10^{-1}$  &  1.06\;\;\;\;\;\;\;\;\;\;\,  \\
A4 & 3.0\,\;       & 0.1 & 1410 &  5.3\,\;  &  1.7\,\;  &  3.00$\times10^{+1}$      &  1.11$\times10^{+2}$  &  6.74\;\;\;\;\;\;\;\;\;\;\,  &  7.82\;\;\;\;\;\;\;\;\;\;\,  \\
A5 &10.0\,\;\;\,  & 0.1 & 2500 &  7.0\,\;  &  2.4\,\;  &2.47$\times10^{+2}$        &  9.56$\times10^{+2}$  &  6.69$\times10^{+1}$  &  6.98$\times10^{+1}$  \\
B1 & 0.10          & 0.2 & 590 &  2.6\,\;  & 0.53      &    1.28$\times10^{-1}$  & 2.69$\times10^{-1}$  & 2.71$\times10^{-2}$ & 3.31$\times10^{-2}$ \\
B2 & 1.0\,\;       & 0.2 & 960 &  5.3\,\;  &  1.2\,\;  &    3.69\;\;\;\;\;\;\;\;\;\;\,   &  1.36$\times10^{+1}$  &  1.42\;\;\;\;\;\;\;\;\;\;\,  &  1.87\;\;\;\;\;\;\;\;\;\;\,  \\
B3 & 3.0\,\;       & 0.2 & 1300 &  6.7\,\;  &  1.6\,\;  &  2.49$\times10^{+1}$      &  9.28$\times10^{+1}$  &  1.16$\times10^{+1}$  &  1.43$\times10^{+1}$  \\
B4 &10.0\,\;\;\,  & 0.2 & 1800 &  7.1\,\;  &  2.1\,\;  &1.62$\times10^{+2}$        &  8.57$\times10^{+2}$  &  1.14$\times10^{+2}$ &  1.49$\times10^{+2}$  \\
C1 & 1.0\,\;       & 0.3 & 900 &  6.1\,\;  &  1.1\,\;  &    2.92\;\;\;\;\;\;\;\;\;\;\,  &  1.09$\times10^{+1}$  &  1.85\;\;\;\;\;\;\;\;\;\;\,  &  2.47\;\;\;\;\;\;\;\;\;\;\,  \\
C2 & 3.0\,\;       & 0.3 & 1200 &  7.8\,\;  &  1.6\,\;  &  2.01$\times10^{+1}$     &  7.42$\times10^{+1}$  &  1.50$\times10^{+1}$  &  1.91$\times10^{+1}$  \\
C3 &10.0\,\;\;\,  & 0.3 & 1650 & 8.1\,\;  &  1.9\,\;  &1.22$\times10^{+2}$       &  6.97$\times10^{+2}$  &  1.57$\times10^{+2}$  &  2.12$\times10^{+2}$  \\
D1 & 0.21          & 0.0 & 1500  & 0.73      &  -        &    3.47$\times10^{-1}$   &  1.37\;\;\;\;\;\;\;\;\;\;\,  &  -    &  -     \\
D2 & 0.70          & 0.0  & 1500 &  1.0\,\;  &  -        &    2.49\;\;\;\;\;\;\;\;\;\;\,   &  1.09$\times10^{+1}$  &  -    &  -     \\
D3 & 2.1\,\;       & 0.0  & 1500 & 1.4\,\;  &  -        &  1.36$\times10^{+1}$      &  8.82$\times10^{+1}$  &  -    &  -     \\
D4 & 7.0\,\;       & 0.0  & 1500 & 1.9\,\;  &  -        &1.06$\times10^{+2}$ &  7.03$\times10^{+2}$  &  -    &  -     \\
\enddata
\end{deluxetable*}

\section{Results} \label{sec:results}

\subsection{Parameter studies}
In this subsection, we discuss the parameter studies of impact simulations. Below, we summarize the adopted parameter survey.
\begin{itemize}
\item We consider the impactor-to-target mass ratio 1, 1/3, and 1/10.
\item We consider two kinds of impactors' atmospheric mass fractions, one being the same as the target's atmospheric mass fraction and the other with no atmosphere.
\item The impact velocity is 1.0-2.6 times escape velocity.
\item We consider only a head-on collision.
\end{itemize}
We summarize all the parameters in Table~\ref{impact_results}.

We compare the impact simulation for the difference the impactor to target mass ratio and the presence of the impactor's atmosphere.
Those comparisons aim to understand the atmosphere loss behavior to consider the energy budget in our discussion.
We demonstrate the impact simulation results 
with the following simulations as examples:
\begin{enumerate}
\item The target is $10~M_\oplus$ with 30\% of the atmosphere by mass, and the impactor is $2.1~M_\oplus$ bare rock with $1.0\times v_\mathrm{esc}$ (Run~31 C3+D3).
\item The target is $1~M_\oplus$ with 30\% of the atmosphere by mass, and the impactor is $0.21~M_\oplus$ bare rock with $1.0\times v_\mathrm{esc}$ (Run~28 C1+D1).
\item The target is $1~M_\oplus$ with 20\% of the atmosphere by mass. The impactor is the same as the target with $1.0\times v_\mathrm{esc}$ (Run~3 B2+B2).
\item The target is $10~M_\oplus$ with 20\% of the atmosphere by mass. The impactor is $1~M_\oplus$ with 20\% of the atmosphere by mass with $1.0\times v_\mathrm{esc}$ (Run~6 B4+B2).
\end{enumerate}
The four kinds of impact simulations mentioned above are categorized into two types:
1 and 2 are the conditions for the impactor without an atmosphere shown in subsection \ref{exatm}.
3 and 4 are the conditions for the impactor with an atmosphere shown in subsection \ref{exrock}.

\subsection{Results for an impactor without an atmosphere} \label{exatm}
In this subsection, we demonstrate the time evolution of the escaping mass when an impactor without an atmosphere collides with the target.

\begin{figure}[htbp]
\begin{center}
\includegraphics[width=\linewidth]{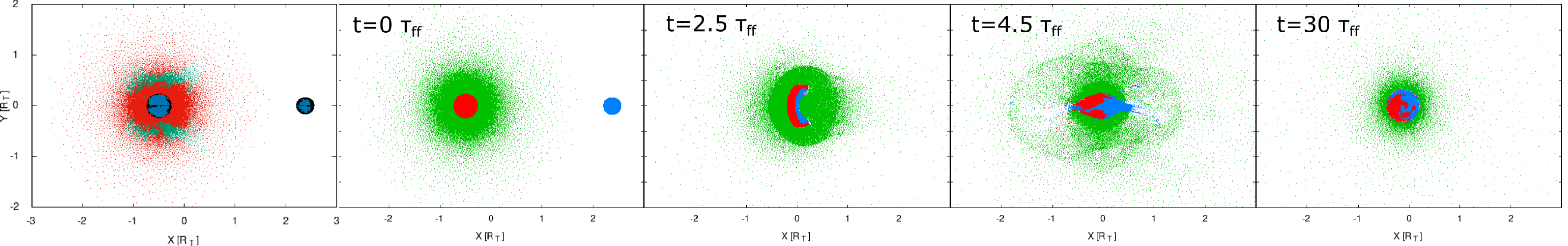}
\includegraphics[width=\linewidth]{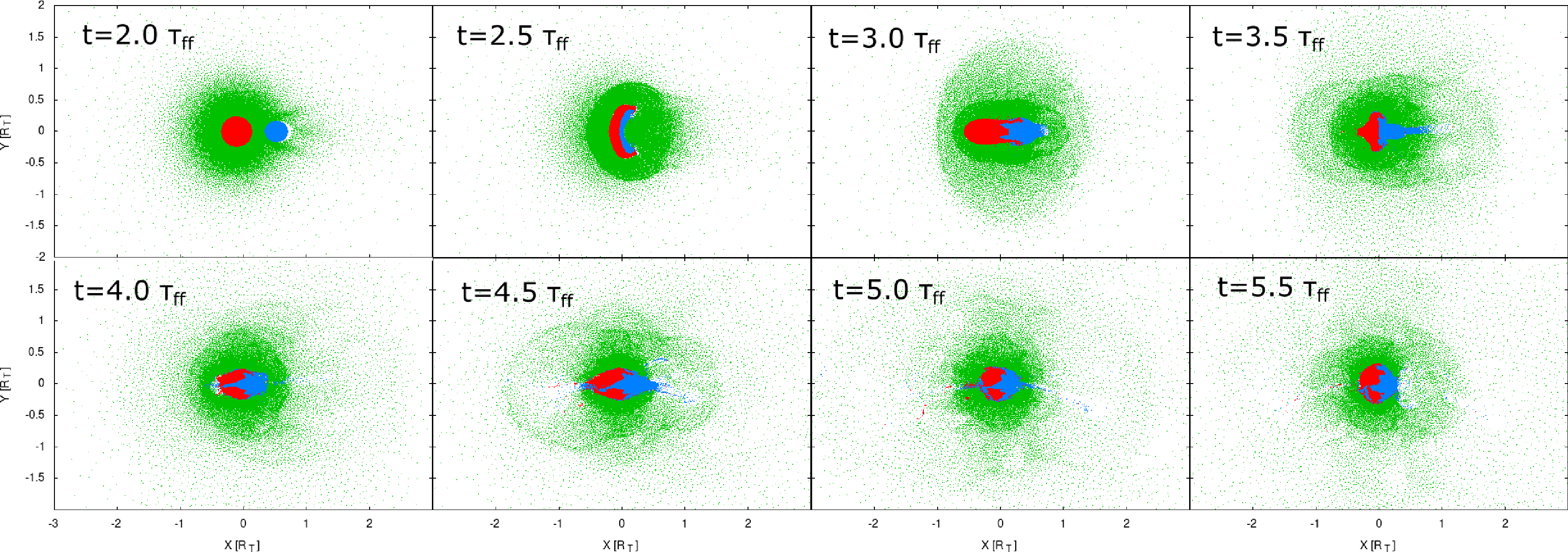}
\caption{
Snapshots of impact simulations.
The left-most panels are color-coded diagrams of lost and bound particles. Red indicates lost atmospheric particles. Green indicates bound atmospheric particles. Black indicates escaping rock particles, and blue indicates bound rock particles.
The top five panels contain simulation snapshots for the target mass as $10~M_\oplus$ with 30\% of the atmosphere by mass, while the impactor mass is $2.1~M_\oplus$ without the atmosphere (Run~31 C3+D3). 
Those snapshots shows the time when $t=0\times\tau_\mathrm{ff}$ (Panel 1), $t=2.5\times\tau_\mathrm{ff}$ (Panel 2), $t=4.5\times\tau_\mathrm{ff}$ (Panel 3), and $t=30\times\tau_\mathrm{ff}$ (Panel 4) from left to right, respectively.
Run~31 simulation is start from $t=-2\times\tau_\mathrm{ff}$, where $\tau_\mathrm{ff}=1.8~$hours.\\
The bottom eight panels shows detail simulation snapshots for Run~31 C3+D3. 
Those snapshots shows the time when $t=2.0\times\tau_\mathrm{ff}$, $t=2.5\times\tau_\mathrm{ff}$, $t=3.0\times\tau_\mathrm{ff}$, 
$t=3.5\times\tau_\mathrm{ff}$, $t=4.0\times\tau_\mathrm{ff}$, 
$t=4.5\times\tau_\mathrm{ff}$, $t=5.0\times\tau_\mathrm{ff}$, and $t=5.5\times\tau_\mathrm{ff}$ from left top to right bottom, respectively.
Colors represent the target and the impactor compositions: Green represents the atmosphere particles of the target and the impactor. Red and blue are the core particles of the target and the impactor, respectively.
\label{12figexample}}
\end{center}
\end{figure}

\begin{figure}[htbp]
\begin{center}
\includegraphics[width=7cm]{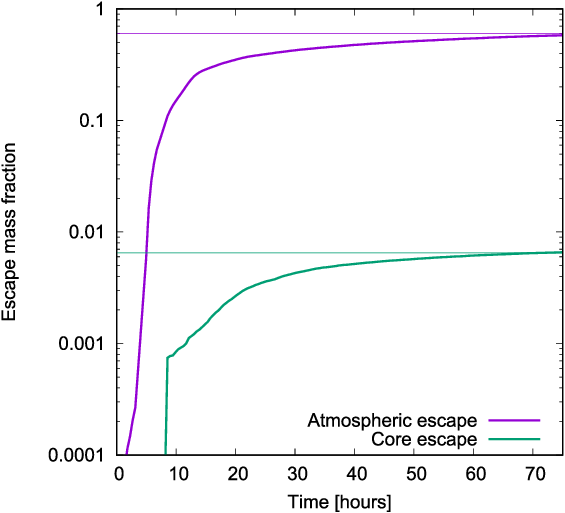}
\includegraphics[width=7cm]{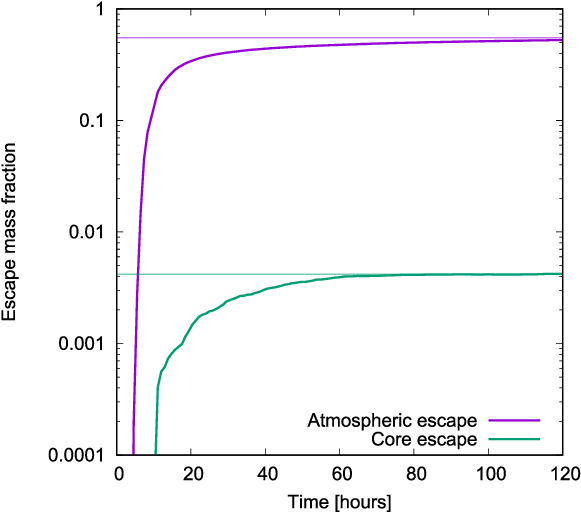}
\caption{Time evolution of the escaping mass fractions for the atmosphere and core.
The left panel shows that the target mass is 10~$M_\oplus$ with 30\% of the atmosphere by mass. The impactor mass is 2.1~$M_\oplus$ without the atmosphere.
The right panel shows the target mass as $1~M_\oplus$ with 30\% of the atmosphere by mass, while the impactor mass is $0.21~M_\oplus$ without the atmosphere. 
The purple line is the atmosphere loss fraction described by Eq~\ref{Hloss}, and the green line is the core loss fraction described by Eq~\ref{Rloss}. 
\label{12lossexample}}
\end{center}
\end{figure}

We choose the target as $10~M_\oplus$ with 30\% of the atmosphere by mass, while the impactor is $2.1~M_\oplus$ without the atmosphere. The target and the impactor collide by 1.0 escape velocity, shown in Run~31.
The top five panels of figure~\ref{12figexample} show snapshots during the impact.
Figure \ref{12figexample} also shows the feature of the cores after the collision for each at $t=2.0\times\tau_\mathrm{ff}$, $t=2.5\times\tau_\mathrm{ff}$, $t=3.0\times\tau_\mathrm{ff}$, $t=3.5\times\tau_\mathrm{ff}$, $t=4.0\times\tau_\mathrm{ff}$, $t=4.5\times\tau_\mathrm{ff}$, $t=5.0\times\tau_\mathrm{ff}$, and $t=5.5\times\tau_\mathrm{ff}$, for Run 31.
After the cores merged, the merged core was observed to extend in the y-direction or the x-direction for every $\sim 0.5 \tau_\mathrm{ff}$.
This non-spherically symmetric oscillation continued, at least, for $5\tau_\mathrm{ff}$ after the collision.
It was also found that the atmosphere gradually lost during this oscillation. 
Therefore, the non-spherically symmetric oscillation after the cores merge is also important in the atmospheric loss discussion.
With time, core particles fall back onto the merged core while the atmosphere expands.
We also find that the deep atmosphere is lost while much of the outer atmosphere remains gravitationally bound because of the atmospheric circulation flow. We explain the typical escaping flow observed in the x- and y-directions. Two major types of outflows are observed in the x-axis direction: One is the impact points flow, and the other is the opposite point of the impact. These flows of the x-axis direction eject the atmosphere. 
In the y-axis direction, the atmosphere is compressed at $t=2.0 \tau_\mathrm{ff}$, but the atmospheric escape occurs in the outermost region of the atmosphere while the inner atmosphere remains. At this moment, the atmosphere around the merged core will flow toward the impact point and the opposite point of impact because the pressure is lower there due to the removal of the atmosphere at the deep, strongly heated atmosphere around the core by the x-direction outflows.
Hence, the atmosphere around the core is expected to flow out by the second x-directional motion of the core starting at $t=4.0 \tau_\mathrm{ff}$ as it moves to a position where it can receive energy from the core's motion again. On the other hand, the outer atmosphere is expected to remain because it will flow around the merged core after the ground motion of the core has weakened.
The left panel of figure~\ref{12lossexample} shows the time evolution of the escaping mass fractions for the atmosphere and core. 
The atmosphere and core loss fraction become constant after the impact reaches 30 free-fall times. We find that the atmosphere loss fraction is 60\% in this simulation result, while the core loss fraction is 0.65\%.
The planetary mass becomes $10.2~M_\oplus$.

\begin{figure}[htbp]
\begin{center}
\includegraphics[width=\linewidth]{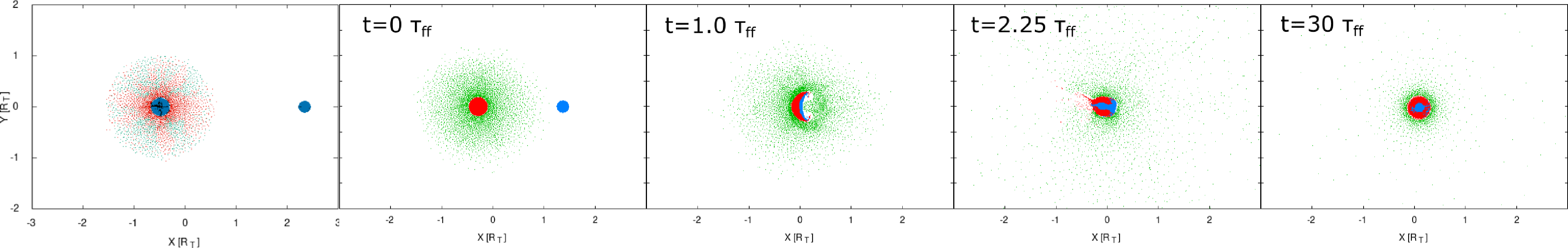}
\caption{Snapshots of impact simulations.
The left-most panels are color-coded diagrams of lost and bound particles. Red indicates lost atmospheric particles. Green indicates bound atmospheric particles. Black indicates escaping rock particles, and blue indicates bound rock particles.
The other four panels contain simulation central cross sections showing the target mass as $1~M_\oplus$ with 30\% of the atmosphere by mass, while the impactor mass is $0.21~M_\oplus$ without the atmosphere (Run~28 C1+D1).
Those snapshots shows the time when $t=0\times\tau_\mathrm{ff}$ (Panel 1), $t=1\times\tau_\mathrm{ff}$ (Panel 2), $t=2.25\times\tau_\mathrm{ff}$ (Panel 3), and $t=30\times\tau_\mathrm{ff}$ (Panel 4) from left to right, respectively.
The colors are the same as in Figure~\ref{12figexample}.
\label{12figexample-2}}
\end{center}
\end{figure}

Next, we consider the second example simulation, with a target whose mass is $1~M_\oplus$ with 30\% of the atmosphere by mass,
and the impactor whose mass is $0.21~M_\oplus$ without an atmosphere colliding by 1.0 escape velocity is shown in Run~28.
The five panels of figure~\ref{12figexample-2} show snapshots during the impact.
The right panel of figure~\ref{12lossexample} shows the time evolution of the escaping mass fraction for the atmosphere and the core. 
Before the impactor's core collides with the target's core; some of the atmosphere is ejected by the shock wave created by the contact of the impactor.
When the impactor collides with the target's core, they merge into a single core.
The merged core underwent a non-spherically symmetric oscillation, and its motion gets weaker as time passes, though the non-spherically symmetric oscillation decays at $t=2.25~\tau_\mathrm{ff}$ that is faster than than Run~31.
As a result, energy is distributed between the core and the atmosphere, causing the atmosphere to expand.
Then the target's atmosphere begins escaping from the system.
The masses of the atmosphere and the core become almost constant after 30 free fall time.
We find that the atmosphere loss fraction is 55\% in this simulation result, while the core loss fraction is 0.42\%.
This result indicates that the escaping materials are dominated by the atmosphere without significantly losing the core.
Thus, the impact event shows that the core mass increases while the atmospheric mass fraction decreases to 13\%.
Finally, the mass of the gravitationally bound body after the impact becomes $1.04~M_\oplus$.

\begin{figure}[htbp]
\begin{center}
\includegraphics[width=8cm]{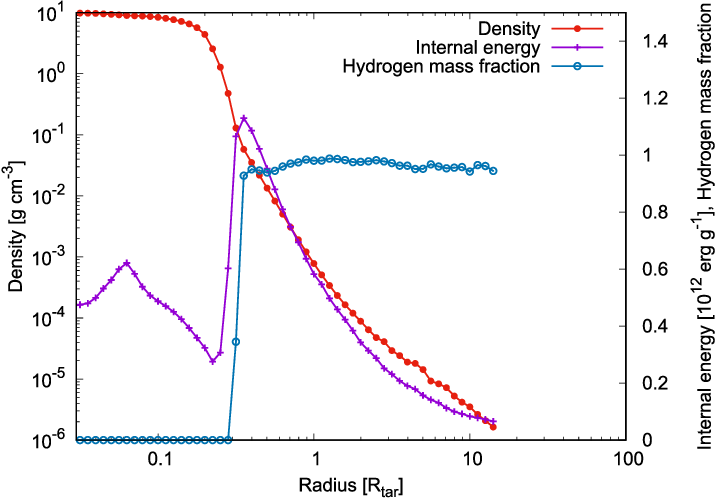}
\includegraphics[width=8cm]{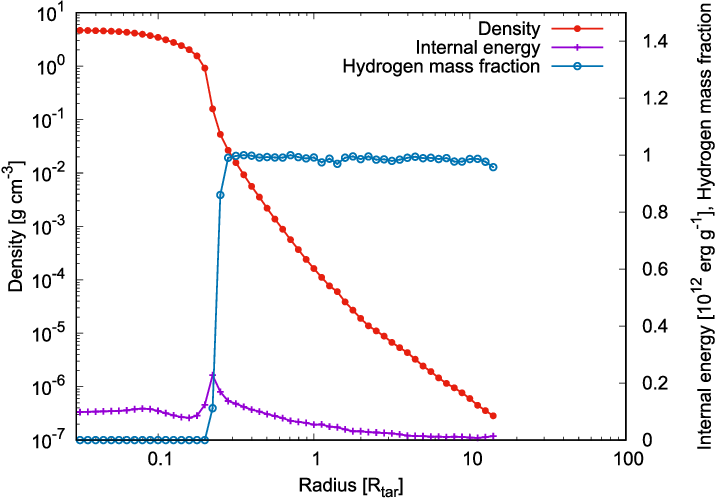}
\caption{The interior structure at $t = 50~\tau_\mathrm{ff}$. 
The radius is normalized by the target radius. 
The left panel shows the target mass as 10~$M_\oplus$ with 30\% of the atmosphere by mass. The impactor mass is 2.1~$M_\oplus$ without the atmosphere. 
The right panel shows the target whose mass is $1~M_\oplus$ with 30\% of the atmosphere, while the impactor whose mass is $0.21~M_\oplus$ without the atmosphere. 
The red line represents the density structure normalized by the central density. 
The purple line reflects the internal energy normalized by $10^{12}~\mathrm{erg}~\mathrm{g}^{-1}$.
The blue line indicates the hydrogen mass fraction. 
\label{12interior}}
\end{center}
\end{figure}

Figure~\ref{12interior} shows the interior structure at $t=50~\tau_\mathrm{ff}$.
We also find that the density and internal energy change steeply at the core and atmosphere interface because the composition changes.
Following the impact, the core oscillation dredges up the core particles in the atmosphere.
With time, core particles that are gravitationally bound fall onto the merged core while the planetary atmosphere expands. 
Since figure~\ref{12interior} shows only gravitationally bound particles, we find that the atmosphere expands up to $\sim 20 R_\mathrm{T}$.
Thus, the planetary interior remains a layered structure.

\subsection{Examples of the numerical results of the impactor with an atmosphere} \label{exrock}
\begin{figure}[htbp]
\begin{center}
\includegraphics[width=\linewidth]{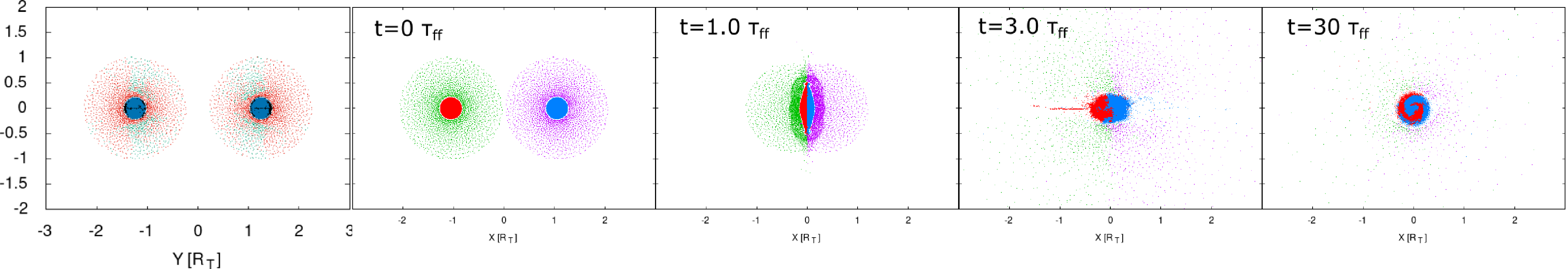}
\includegraphics[width=\linewidth]{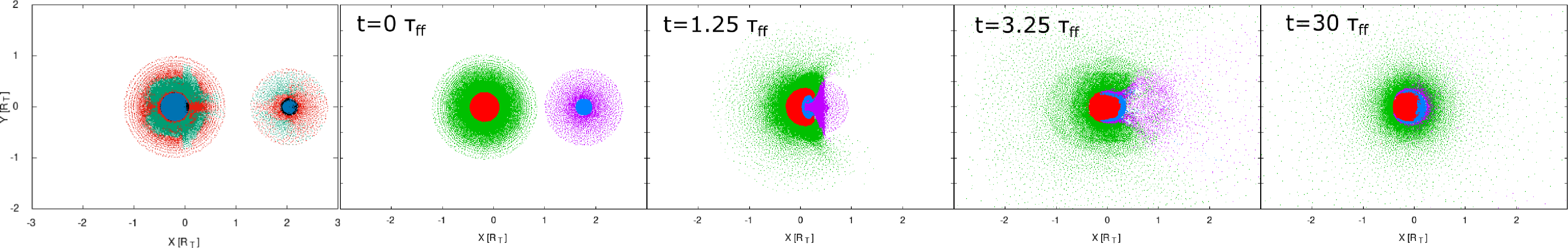}
\caption{
Snapshots of impact simulations. 
Top five panels show simulation central cross sections reflecting the target mass as $1~M_\oplus$ with 20\% of the atmosphere by mass, while the impactor mass is $1~M_\oplus$ with 20\% of the atmosphere by mass (Run~3 B2+B2).
The top left-most panel is color-coded diagrams of lost and bound particles.
Red indicates lost atmospheric particles. Green indicates bound atmospheric particles. Black indicates escaping rock particles, and blue indicates bound rock particles.
Those snapshots shows the time when $t=0\times\tau_\mathrm{ff}$ (Panel 1), $t=1\times\tau_\mathrm{ff}$ (Panel 2), $t=3\times\tau_\mathrm{ff}$ (Panel 3), and $t=30\times\tau_\mathrm{ff}$ (Panel 4) from left to right, respectively.
Run~3 simulation is start from $t=-1\times\tau_\mathrm{ff}$, where $\tau_\mathrm{ff}=3.0~$hours. 
\\
The bottom five panels show simulation snapshots indicating the target mass as $10~M_\oplus$ with 20\% of the atmosphere by mass, and impactor mass is $1~M_\oplus$ with 20\% of the atmosphere by mass (Run~6 B4+B2).
The bottom left-most panel is color-coded diagrams of lost and bound particles, whose colors is the same as the top left-most panel.
The other bottom four snapshots shows the time when $t=0\times\tau_\mathrm{ff}$ (Panel 1), $t=1.25\times\tau_\mathrm{ff}$ (Panel 2), $t=3.25\times\tau_\mathrm{ff}$ (Panel 3), and $t=30\times\tau_\mathrm{ff}$ (Panel 4) from left to right, respectively.
Run~6 simulation is start from $t=-1\times\tau_\mathrm{ff}$, where $\tau_\mathrm{ff}=1.5~$hours.
The colors are the same as in Figure~\ref{12figexample}.
 \label{34figexample}}
\end{center}
\end{figure}

\begin{figure}[htbp]
\begin{center}
\includegraphics[width=7cm]{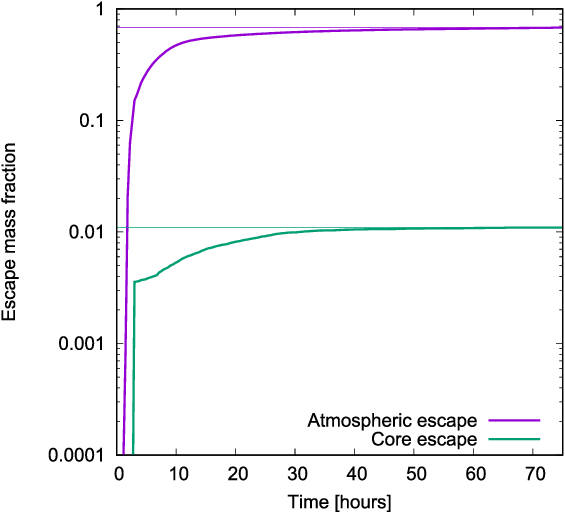}
\includegraphics[width=7cm]{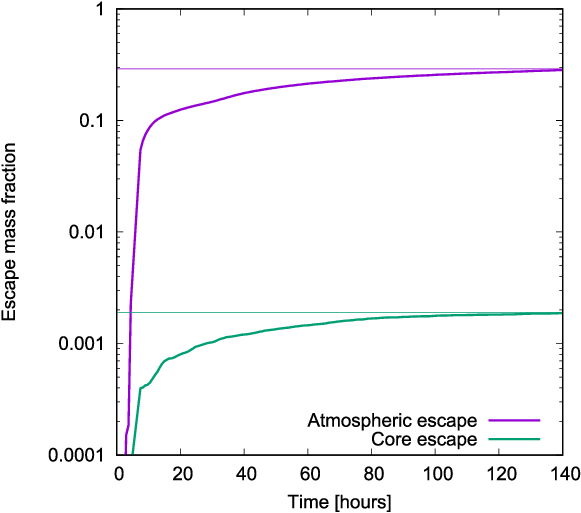}
\caption{Time evolution of the escaping mass fraction after the impact event.
The left panel shows the target whose mass is $1~M_\oplus$ with 20\% of the atmosphere by mass while the impactor is the same as the target.
The right panel shows the target whose mass is $10~M_\oplus$ with 20\% of the atmosphere by mass while the impactor whose mass of $1~M_\oplus$ with 20\% of the atmosphere by mass.
Colored lines are same as those in Figure~\ref{12lossexample}.
\label{34lossexample}}
\end{center}
\end{figure}

In this subsection, we consider the time evolution of the escaping mass for the impactor with an atmosphere.
We show the target mass as $1~M_\oplus$ with 20\% of the atmosphere by mass, and the impactor is the same as the target colliding by 1.26 escape velocity shown in Run~3.
The five panels of figure~\ref{34figexample} show snapshots during the impact.
The left panel of figure~\ref{34lossexample} shows the time evolution of the escaping mass fraction for the atmosphere and core. 
Panel 2 in the top panels of figure~\ref{34figexample} shows the impactor's core collides with the target's core.
Target's and impactor's atmospheres are compressed before their cores collide because of the interaction between their atmospheres.
When the impactor collides with the target's atmosphere, the target's atmosphere ejects from the impact point.
The atmosphere is ejected from the antipodes when the impactor's core collides with the target's core. The target's atmosphere begins to eject during the oscillation of the merged core. The slightly spiral pattern in the merged core is due to the fluctuation of the initial particle configuration of the target's and impactor's core.
We find that the atmosphere loss fraction is 68\%, while the core loss fraction is 1.1\%.
As a result of the impact event, the merged planetary mass becomes 1.71 Earth-mass, while the atmosphere mass fraction decreases to 7.5\%

\begin{figure}[htbp]
\begin{center}
\includegraphics[width=8cm]{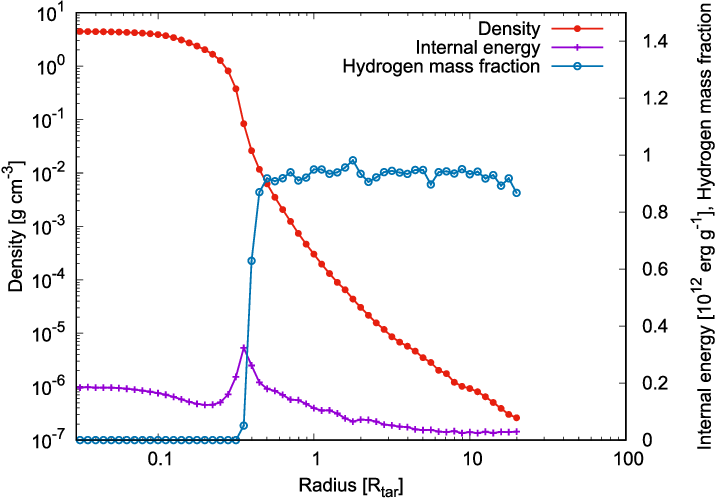}
\includegraphics[width=8cm]{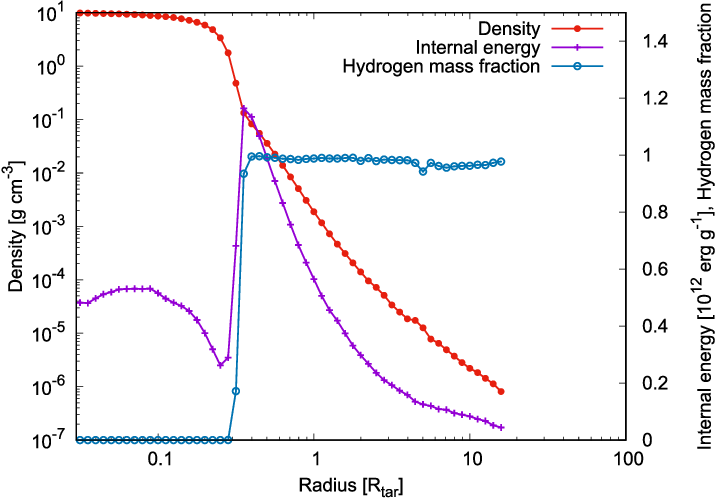}
\caption{The interior structure at $t = 50~\tau_\mathrm{ff}$. 
The radius is normalized by the target radius. 
The left panel shows the target whose mass is $1~M_\oplus$ with 20\% of the atmosphere by mass while the impactor is the same as the target.
The right panel shows the target whose mass is $10~M_\oplus$ with 20\% of the atmosphere by mass while the impactor whose mass of $1~M_\oplus$ with 20\% of the atmosphere by mass.
Colored lines are the same as those in Figure~\ref{12interior}.
\label{34interior}}
\end{center}
\end{figure}

Next, we consider the target whose mass is $10~M_\oplus$ with 20\% of the atmosphere and the impactor whose mass is $1~M_\oplus$ with 20\% of the atmosphere colliding by 1.65 escape velocity, which is shown in Run~6.
The bottom five panels of figure~\ref{34figexample} show snapshots during the impact.
When the impactor's core collides with the target's core, the impactor makes a large crater on the target. The shock from the target to the impactor pushes the impactor's core. Then core particles at the surface of the impactor are mixed into the atmosphere. After that, the impactor's core materials fall onto the target's core and cover the surface of the target. That is why the blue particle covers the red particles.
The right panel of figure~\ref{34lossexample} shows the time evolution of the escaping mass fraction of the atmosphere and core.
When the impactor's core collides with the target's (see Panel 2 in the bottom panels of figure \ref{34figexample}), the impactor's atmosphere is compressed by the target's atmosphere. 
During the impact, the impactor's atmosphere expands from the antipodes of the impact point
while the target's and impactor's cores merge into one, oscillating after the merger. 
This simulation result showed that the atmosphere loss fraction is 29\%, while the core loss fraction is 0.2\%.
In this case, the atmospheric mass fraction after the impact becomes 15\%.
The impact increases the planetary mass to 10.3 Earth masses.
Figure~\ref{34interior} shows the interior structure at $t=50~\tau_\mathrm{ff}$.
In this impact case, the impact energy is relatively small, and the planetary interior structure remains layered, similar to the figure~\ref{12interior}.

\section{Analysis of the energy budget} \label{sec:anal}
In this section, we analyze the energy budget before and after the collision to study the kinetic energy of the escaping particles and the impact-induced atmosphere loss.
We first summarize the known and unknown parameters.
We know the kinetic impact energy $K_\mathrm{imp}$, the pre-impact structure for the target $\Phi_T$ and $U_T$, and the impactor $\Phi_I$ and $U_I$. We assume the core is merged completely. This section derives the unknown value for the escaping mass $M_\mathrm{esc}$.
Once we find $M_\mathrm{esc}$,
the post-impact planetary mass and the atmospheric mass fraction are determined by Eq.~\ref{Mb-def} and \ref{Yb-def}, respectively. 
To derive $M_\mathrm{esc}$,
we consider the kinetic energy of escaping particles $K_\mathrm{esc}$, the gravitational energy and internal energy of the post-impact planet $\Phi_B$ and $U_B$. Since the total energy is conserved before and after the impact, $K_\mathrm{esc}$ should be described by the energy conservation law.
This section describes the energy distribution between the kinetic energy of the escaping atmosphere, the bound energy of the atmosphere, and the internal energy of the core before and after the impact. There are three kinds of parameters to describe the distribution, the atmospheric bound energy ratio before and after the impact $(\varepsilon)$, the correlation between the kinetic energy of escaping atmosphere and the post-impact planet's escape velocity $(\beta)$, and the ratio of the kinetic energy of the escaping atmosphere to the heating of the rock core $(\alpha)$.
First, we take five steps to consider the impact-induced atmosphere loss.
We examine the SPH simulation for energy coservation
before and after the collision (section \ref{sec:eng}). Second, we determine the relationship between the kinetic energy of the escaping particles and the atmosphere mass loss (section \ref{sec42}). Third, we consider the energy release induced by the merged core (section \ref{sec43}). Fourth, we consider the energy distribution between the kinetic energy of escaping mass, the atmosphere's total energy, and the core's internal energy (section \ref{sec44}, \ref{sec45}). Finally, we summarize the analytic prediction for the atmosphere loss fraction (section \ref{sec46}).

\subsection{The energy conservation before and after the impact} \label{sec:eng}
Before we examine where the impact energy is distributed during the impact, we must first check the conservation property of our SPH simulations.
The suffixes T, I, and B indicate the target, impactor, and gravitationally bound body after the collision.
We denote the target's total internal energy and total gravitational energy as $U_\mathrm{T} = U_\mathrm{T,atm}+U_\mathrm{T,c}$ and $\Phi_\mathrm{T} = \Phi_\mathrm{T,atm}+\Phi_\mathrm{T,c}$.
Similarly, we denote the the impactor's total internal energy and total gravitational energy as $U_\mathrm{I} = U_\mathrm{I,atm}+U_\mathrm{I,c}$ and $\Phi_\mathrm{I} = \Phi_\mathrm{I,atm}+\Phi_\mathrm{I,c}$, while the gravitationally bound body's total internal energy and total gravitational energy are $U_\mathrm{B} = U_\mathrm{B,atm}+U_\mathrm{B,c}$ and $\Phi_\mathrm{B} = \Phi_\mathrm{B,atm}+\Phi_\mathrm{B,c}$.
We denote $\Phi_\mathrm{TI}$ as the mutual gravitational energy between the target and impactor.
The kinetic impact energy $K_\mathrm{imp}$ should include $\Phi_\mathrm{TI}$ because the mutual gravitational energy increases the target's and impactor's velocities.

\begin{figure}[htbp]
\begin{center}
\includegraphics[width=\linewidth]{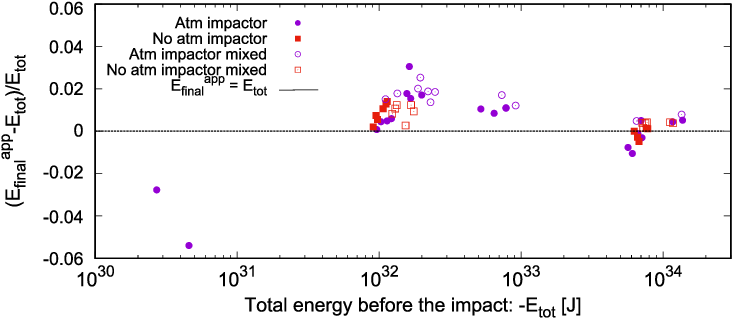}
\caption{The relationship between the initial total energy calculated by Eq.~\ref{Etot} and the approximate total energy after a collision calculated by the right-hand side of Eq.~\ref{Kesc-anal}.
Purple circles are the impact simulations when the impactor has an atmosphere, while red squares are the bare-rock impactor.
Filled marks reflect that the planet retains a layered structure after the impact, while open marks represent the planet has a region wherein the atmosphere and core are mixed.
The black dotted line shows $E_\mathrm{final}^\mathrm{app}=E_\mathrm{tot}$.
\label{Kesc_hikaku}}
\end{center}
\end{figure}

Since the total energy is conserved during the collision, we have the following relation:
\begin{equation}
E_\mathrm{tot} =K_\mathrm{imp} + U_\mathrm{T} + \Phi_\mathrm{T} + U_\mathrm{I} + \Phi_\mathrm{I}. \label{Etot}
\end{equation}
where $K_\mathrm{imp}$ is the kinetic impact energy shown as $K_\mathrm{imp}=\frac{1}{2}\mu v_\mathrm{rel}^2 +\Phi_\mathrm{TI}$.
The total linear momentum of the escaping material is not exactly zero but negligible. We ignore it and assume that the resultant planet after the collision remains in the center of mass of the system. The approximate total energy after the collision can be written as
\begin{equation}
E_\mathrm{final}^\mathrm{app}=  K_\mathrm{esc}  + (U_\mathrm{B} + \Phi_\mathrm{B}). \label{Kesc-anal}
\end{equation}
Note that Eq.~\ref{Kesc-anal} also ignores the gravitational energy of escaping particles.
After a few times of the non-spherically symmetric oscillations of the resulting object, the kinetic energy of the bound material becomes negligibly smaller than its gravitational energy so that we ignore the former.
Figure~\ref{Kesc_hikaku} shows the relationship between  
the total energy $E_\mathrm{tot}$ (Eq.~\ref{Etot}) and $E_\mathrm{final}^\mathrm{app}$ (Eq.~\ref{Kesc-anal}).
Since the internal energy or gravitational energy of escaping particles is 0.1\% of their kinetic energy,
we can estimate the energy of the escaping 
material as $K_\mathrm{esc} = E_\mathrm{tot}-(U_\mathrm{B}+\Phi_\mathrm{B})$.
Figure~\ref{Kesc_hikaku} plots all the cases for the impactors with and without atmosphere and thus we find that approximate expression
shown in Eq.~\ref{Kesc-anal} is valid regardless of whether the impactor has an atmosphere.
The error in the energy conservation before and after the impact is at most 6\% by Eq.~\ref{Kesc-anal}.

When the atmosphere loss is significant, we observe that the atmosphere gas and core material are mixed after the impact because parts of the core material are dredged up after the impact (see also section~\ref{miximpact}). This is significant in the case for planets whose mass is 1 Earth-mass with 10 \% of atmosphere. However, figure~\ref{Kesc_hikaku} shows almost equivalence between the total energy and approximate final energy, which means that Eq.~\ref{Kesc-anal} is valid for the cases of both layered structures and mixed structures after the collision. Thus, we can accurately estimate  $K_\mathrm{esc}$ by $E_\mathrm{tot}-(U_\mathrm{B}+\Phi_\mathrm{B})$.

\subsection{The mass of escaping atmosphere} \label{sec42}
In this subsection, we estimate the escaping mass using the kinetic energy of the escaping particles.
The bound mass $M_\mathrm{B}$ is $M_\mathrm{T}+M_\mathrm{I}-M_\mathrm{esc}$ such that $M_\mathrm{B}$ includes the known values ($M_\mathrm{T}$ and $M_\mathrm{I}$) and the unknown value ($M_\mathrm{esc}$). In this subsection, both $M_\mathrm{esc}$ and $K_\mathrm{esc}$ are unknown parameters, while $K_\mathrm{esc}$ is represented by known variables shown in the next sections. We can make the relationship between $K_\mathrm{esc}$ and $M_\mathrm{esc}$ when the escaping particles' mean velocity is predicted.
This section aims to determine the correlation between $K_\mathrm{esc}$ and $M_\mathrm{esc}$. Given the $M_\mathrm{esc}$, the post-impact planetary mass $(M_B)$ and the atmosphere mass fraction $(Y_B)$ are deduced because most impact-induced mass loss originates from the atmosphere.

We assume that the typical velocity of escaping masses is the escape velocity at the surface of the $M_B$,
whose radius ($R_B$) is estimated using $M_B$ and $Y_B$. $R_B$ is estimated by the interpolation from Table~\ref{planetary_model}, assuming that the radius of the gravitationally bound body is a function of the planetary mass and the atmosphere mass fraction.
Thus, $K_\mathrm{esc}$ should be represented by $M_\mathrm{esc}$, $M_B$, $Y_B$, and $R_B(M_B, Y_B)$.
Since the planetary radius is complex function about the mass and atmospheric mass, we use mean density instead of the radius. We find that the mean density is monotonic increase by mass and core mass in our model. Thus, the planet's density is used for determining the escape velocity at the surface.
To predict the planetary radius after the impact, we make a regression line between the planetary mass, the atmospheric mass fraction, and the planetary radius from Table~\ref{planetary_model} and find
\begin{eqnarray}
R(M_B,Y_B) &=& \left(\frac{3M_B}{4\pi \rho(M_B,Y_B)} \right)^{1/3} \label{Rb-def} \\
\log_{10}\left(\frac{\rho(M_B,Y_B)}{\rho_0} \right) &=& a_1 + a_2 \log_{10}\left( \frac{M_B}{M_\oplus} \right)  
+ a_3 Y_B + a_4 \log_{10}\left( \frac{M_B}{M_\oplus} \right) Y_B  + a_5 \left[\log_{10}\left( \frac{M_B}{M_\oplus} \right) \right]^2 + a_6 Y_B^2 \label{Rhob-def}
\end{eqnarray}
where $\rho(M_B,Y_B)$ is the bulk density, $\rho_0$ is the typical density for the bare-rock planet, which is determined by the fitting of the Table~\ref{planetary_model} of the impactor.
We find that $\rho_0 = 3.0~\mathrm{g}\cdot\mathrm{cm}^{-3}$,
 $a_1=2.0\times10^{-3}$, $a_2=0.19$, $a_3=-18.0$, $a_4= 0.52$, $a_5=0.16$, and $a_6=38.2$.
In this study, we define $R_B = R(M_B,Y_B)$ and $\rho_B = \rho(M_B,Y_B) $ for simplicity.
Eq.~\ref{Rhob-def} is valid for $0.1 \le M_B/M_\oplus \le 10$ and $0 \le Y_B \le 0.3$.
We estimate the relationship between the $K_\mathrm{esc}$ and the escape velocity determined by the surface of the gravitationally bound body.
We set the escape velocity determined by the surface of the gravitationally bound mass $v_\mathrm{B,esc}(=\sqrt{2GM_B/R_B}$).
We define $E_\mathrm{esc,ref}$ as the reference kinetic energy for gravitational escape velocity when the velocities of all the materials are assumed to be $v_\mathrm{B,esc}$ which is shown as
\begin{equation}
E_\mathrm{esc,ref} = \frac{GM_BM_\mathrm{esc}}{R_B} \label{Esc-bound1}
\end{equation}
where $M_\mathrm{esc}$ is the escaping mass, $M_B$ is the mass of the gravitationally bound body,  
and $R_B$ is the radius of the gravitationally bound body.
If the velocity of all escaping particles ($M_\mathrm{esc}$) is $v_\mathrm{B,esc}$, the kinetic energy of escaping particles is 
$E_\mathrm{esc,ref}$
shown in the right-hand side of Eq.~\ref{Esc-bound1}.
We call the right-hand side of Eq.~\ref{Esc-bound} the kinetic energy of escaping particles at the surface escape velocity of the bound body.
If no excess kinetic energy was imparted to the escaping particles, then we can expect that 
$K_\mathrm{esc} = E_\mathrm{esc,ref}$.
The atmosphere particles are accelerated as the shock wave propagates through the atmosphere. At this time, the atmosphere particles escape when the velocity of those particles at that location exceeds the escape velocity of the atmosphere.
In that case, escaping particles can have velocities smaller than $v_\mathrm{B,esc}$.
We also check the velocity distributions of core and atmosphere particles after the impact.
The velocity distributions of the escaping particles revealed that the velocity of most of the escaping particles is scaled by $v_\mathrm{B,esc}$ (see Appendix~\ref{app-esc}).
Figure~\ref{Reff_simple} shows the relationship between $K_\mathrm{esc}$ and $E_\mathrm{esc,ref}$.
We find that $K_\mathrm{esc}$ is approximated by 
\begin{equation}
K_\mathrm{esc} = \beta E_\mathrm{esc,ref}, \label{Esc-bound}
\end{equation}
where $\beta=0.42$.
This result implies that the mean velocity of escaping particles is slightly slower than the escape velocity at the surface of the bound body because the atmosphere expands, and the atmosphere begins to escape.
Then, the radius that determines the mean escape velocity is larger than $R_B$.

\begin{figure}[htbp]
\begin{center}
\includegraphics[width=\linewidth]{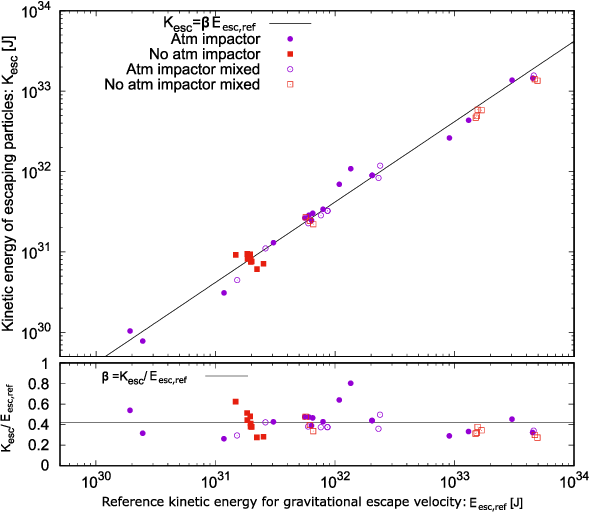}
\caption{
The relationship between the total kinetic energy of escaping particles $(K_\mathrm{esc})$ and the reference kinetic energy for gravitational escape velocity $(E_\mathrm{esc,ref})$ defined by Eq.\ref{Esc-bound1}.
The bottom panel shows the value of $\beta=K_\mathrm{esc}/E_\mathrm{esc,ref}$ for all numerical results. The black line shows a linear regression, $K_\mathrm{esc} = \beta E_\mathrm{esc,ref}$, where $\beta=0.42$.
Plot colors are the same as those in Figure~\ref{Kesc_hikaku}.
\label{Reff_simple}}
\end{center}
\end{figure}

\subsection{Significance of Gravitational Energy} \label{sec43}
In previous section, we have found the correlation between the $K_\mathrm{esc}$ and $M_\mathrm{esc}$ by Eq.~\ref{Esc-bound}. Eq.~\ref{Esc-bound} assumed that the average velocity of the escaping mass is the escape velocity of the bound body.
In this subsection, we analyze the energy budget for the impact to estimate the energy released from the merged core.
Suppose that an impactor falls onto the target from an infinite distance, 
where the target's and impactor's core masses are $M_\mathrm{T,c}$ and $M_\mathrm{I,c}$, respectively.
Some of the impact energy will heat the atmosphere and the core, while the rest of the energy will be used in the atmosphere loss. Since the gravitational potential energy is a function of the planetary mass, a massive planet has large gravitational potential. The energy release induced by the merged core is described by
\begin{equation}
\delta\Phi_{c} = \Phi(M_\mathrm{T,c}+M_\mathrm{I,c},R_\mathrm{B,c}) - \Phi(M_\mathrm{T,c},R_\mathrm{T,c}) - \Phi(M_\mathrm{I,c},R_\mathrm{I,c})
\label{Phic},
\end{equation}
where $R_\mathrm{T,c}, R_\mathrm{I,c}$ and $R_{B,c}$ are the radii of the target, impactor, and merged cores that are gravitationally bound after the impact.
We consider the case where $M_\mathrm{T,c} =M_\mathrm{I,c}$.
We set the core mass $M_c$ and the core's mean density $\rho_c$.
$\rho_c$ does not change significantly before and after the impact. 
The gravitational potential energy is given by $\Phi(M_\mathrm{c}, R_\mathrm{c})=-AGM_\mathrm{c}^2/R_\mathrm{c} (=\Phi_\mathrm{T,c})$ where $A$ is the factor on the order of unity.
\footnote{
When the planetary structure is determined by the polytrope $(P\propto \rho^{\gamma})$, we find $A=3(\gamma-1)/(5\gamma-6)$ \citep[e.g.,][]{Chandrasekhar1967}. According to our numerical results, $A\approx 0.2$ is more precise than the constant density result $(A=3/5)$.
}
In this case, we find
\begin{equation}
\delta\Phi_c = - AG \left( \frac{4\pi\rho_c}{3} \right)^{1/3} \left[ (M_\mathrm{T,c} + M_\mathrm{I,c})^{5/3} - M_\mathrm{T,c}^{5/3} - M_\mathrm{I,c}^{5/3} \right].\label{Eq-delPhic}
\end{equation}
In the case $M_\mathrm{T,c} = M_\mathrm{I,c}$, $\delta\Phi_c = 1.17~\Phi_\mathrm{T,c}$, which is even larger than the total energy of two objects before the collision.
Eq.~\ref{Eq-delPhic} indicates that $\delta \Phi_c$ is positive and very large.
Accordingly, the released energy from the merged core significantly heats the planet.
$\delta\Phi_c$ transforms into the core's and the atmosphere's internal energy.
That is, $K_\mathrm{imp}-\delta \Phi_c$ would be useful to consider the energy contribution between the heating of the atmosphere, the kinetic energy of the escaping particles, and the heating of the core.
In the next section, we examine the distribution for the energy.

\subsection{The contribution of the atmospheric energy} \label{sec44}

\begin{figure}[htbp]
\begin{center}
\includegraphics[width=\linewidth]{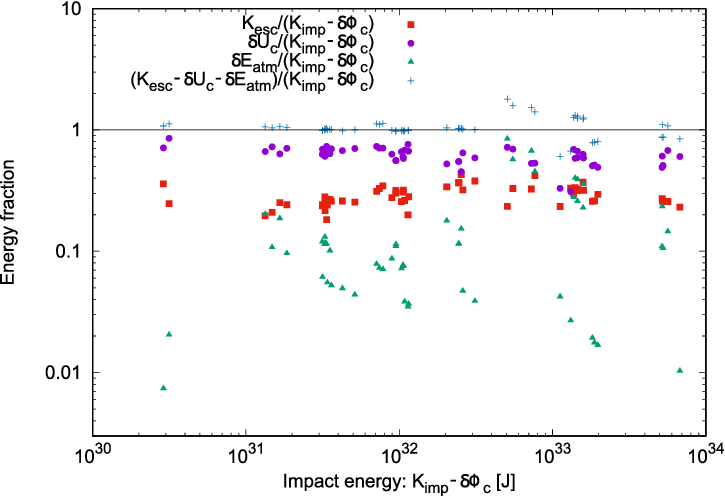}
\caption{The relationship between the sum of the kinetic impact energy and the released energy from the merged core $K_\mathrm{imp}-\delta \Phi_c$, the kinetic energy of escaping mass $K_\mathrm{esc}$, and energy changes $\delta E_\mathrm{atm}, \delta U_c$.
$\delta \Phi_c$ is calculated by Eq.~\ref{Phic}.
Plot colors represent the energy changes. 
Red plots are the kinetic energy of escaping mass (Eq.~\ref{Kesc-def}), purple plots are the change of the internal energy of the core (Eq.~\ref{deU1}), and the green plots are the change of the energy of the atmosphere (Eq.~\ref{deH1}).
Blue plots represent the fraction of 
$K_\mathrm{esc}-\delta U_c-\delta E_\mathrm{atm}$ to $K_\mathrm{imp}-\delta \Phi_c$.
\label{betaestimate}}
\end{center}
\end{figure}

In this subsection, we study the energy distribution between the kinetic energy of escaping mass, the internal energy of the core, and the sum of the gravitational and internal energy of the bound atmosphere.
We denote the atmosphere's total energy of the target, the impactor, and the gravitationally bound body as  $E_\mathrm{T,atm} = U_\mathrm{T,atm}+\Phi_\mathrm{T,atm}$, $E_\mathrm{I,atm} = U_\mathrm{I,atm}+\Phi_\mathrm{I,atm}$, and $E_\mathrm{B,atm} = U_\mathrm{B,atm}+\Phi_\mathrm{B,atm}$, respectively.
We consider the change of the sum of the bound atmosphere's the gravitational and internal energy $\delta E_\mathrm{atm}$ and the core's internal energy $\delta U_c$ shown as
\begin{eqnarray}
\delta E_\mathrm{atm} &=& E_\mathrm{B,atm} - (E_\mathrm{T,atm}+E_\mathrm{I,atm}), \label{deH1}\\
\delta U_c &=& U_\mathrm{B,c} - (U_\mathrm{T,c}+U_\mathrm{I,c}). \label{deU1}
\end{eqnarray}
Figure~\ref{betaestimate} shows the relationship between the sum of the kinetic impact energy and the released energy from the merged core $K_\mathrm{imp}-\delta \Phi_c$, the kinetic energy of escaping mass $K_\mathrm{esc}$, the energy change of the atmosphere $\delta E_\mathrm{atm}$, and the internal energy change of the core $\delta U_c$.
Our results show that $\delta E_\mathrm{atm}$ and $\delta U_c$ are positive because the atmosphere and the core receive the kinetic energy induced by the impact.
For most impacts, about 30 \% of the energy goes to the escaping material, 70 \% to the heating core, and 1--10 \% to the bound atmosphere.
That is, the kinetic impact energy is distributed to the $K_\mathrm{esc}$,  $\delta U_c$, and $\delta E_\mathrm{atm}$.

\begin{figure}[htbp]
\begin{center}
\includegraphics[width=\linewidth]{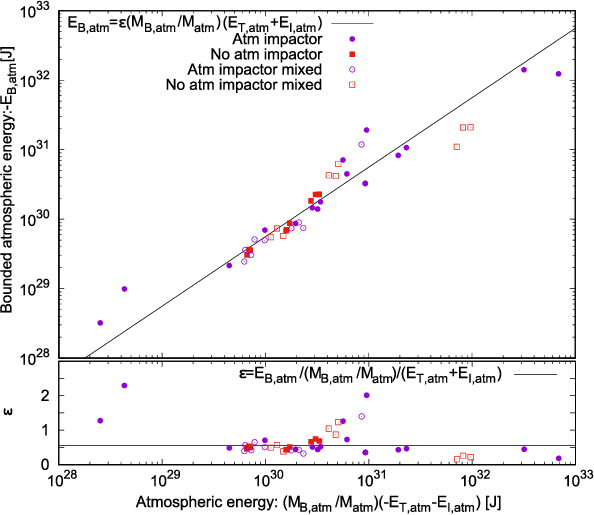}
\caption{
The comparison of the energy of bound atmosphere calculated by Eq.~\ref{H-eng-def} and numerical results $E_\mathrm{B,atm}$.
Black line represent the linear regression result $\epsilon = 0.56$.
Plot colors are the same as those shown in Figure~\ref{Kesc_hikaku}.
\label{PhidelUH}}
\end{center}
\end{figure}

We consider the contribution of atmospheric energy change $\delta E_\mathrm{atm}$ to $K_\mathrm{esc}$.
We assume the relationship of the atmospheric energies before and after the collision as
\begin{equation}
E_\mathrm{B,atm} = \epsilon \frac{M_\mathrm{B,atm}}{M_\mathrm{atm}}(E_\mathrm{T,atm}+E_\mathrm{I,atm}) \label{H-eng-def}
\end{equation}
where $\epsilon$ is a proportionality constant, $M_\mathrm{atm}$ is the total atmospheric mass denoted by $M_\mathrm{atm} = M_\mathrm{T,atm} + M_\mathrm{I,atm}$, and $M_\mathrm{B,atm}$ is the atmospheric mass after the impact.
Since $M_\mathrm{B,atm}/M_\mathrm{atm} = 1-X_\mathrm{atm}$ as defined by Eq.~\ref{Hloss}, we set the atmospheric energy after the impact $E_\mathrm{B,atm}$ is a function of the atmospheric loss fraction $X_\mathrm{atm}$.
Figure~\ref{PhidelUH} plots the atmospheric energy change, which shows that the bounded atmospheric energy seems to be proportional to the initial energy multiplied by the mass fraction of the remaining atmosphere.
We fit the results by Eq.~\ref{H-eng-def} and find $\epsilon = 0.56$.
The planetary radius is expanded after the impact because the atmosphere obtains the impact energy.
Thus, we can represent $K_\mathrm{esc}$ from Eq.~\ref{Kesc-anal}, \ref{deH1}, \ref{deU1}, and \ref{H-eng-def} as
\begin{eqnarray}
K_\mathrm{esc} & = & K_\mathrm{imp} - \delta \Phi_c -\delta E_\mathrm{atm} - \delta U_{c}   \nonumber \\
&=& E_\mathrm{av} - \delta U_c, \label{Kesc-wHatm}
\end{eqnarray}
where $E_\mathrm{av}$ is defined as the avaliable enegy shown as
\begin{equation}
E_\mathrm{av} =  K_\mathrm{imp} - \delta \Phi_c -\delta E_\mathrm{atm}. \label{Eav}
\end{equation}
Then, $K_\mathrm{esc}$ equals the remainder of the impact energy subtracted by the atmospheric expansion and the heating of the core.
Note that $\delta E_\mathrm{atm}$ term is significant only in the case a substantial atmosphere remains after the impact event.
When the planet has no atmosphere after the impact, the available anergy is approximately described as $K_\mathrm{imp} - \delta \Phi_c$.
When the $M_\mathrm{esc}$ is negligible, on the other hand, the available energy is described as $K_\mathrm{imp} - \delta \Phi_c + (1-\varepsilon) (E_\mathrm{T,atm} + E_\mathrm{I,atm})$ which consider the available energy subtracted by the atmospheric expansion energy.

\subsection{The contribution of the core energy} \label{sec45}
As discussed in section \ref{sec44}, the remainder of the impact energy subtracted by the atmospheric expansion is available for $K_\mathrm{esc}$ and the core heating.
Eq.~\ref{Kesc-wHatm} suggests that the relationship between $\delta U_c$ and $E_\mathrm{av}$ is crucial in determining $K_\mathrm{esc}$. 
We derive the equation combined with Eq.~\ref{Kesc-wHatm} and \ref{alpha-def}
\begin{equation}
K_\mathrm{esc}  = \alpha E_\mathrm{av}. \label{KescKinput}
\end{equation}
where $\alpha$ is a proportionality constant defined as 
\begin{equation}
\alpha = 1 - \frac{\delta U_c}{E_\mathrm{av}}. \label{alpha-def}
\end{equation}
First, we obtain the relationship between $\delta U_c$ and $E_\mathrm{av}$ shown in Eq.~\ref{alpha-def} by the linear regression. Second, we obtain the relationship between $K_\mathrm{esc}$ and $E_\mathrm{av}$ shown in Eq.~\ref{KescKinput} by the linear regression. Finally, we check the consistency of the proportionality constant $\alpha$ obtained by Eq.~\ref{alpha-def} and \ref{KescKinput}.

\begin{figure}[htbp]
\begin{center}
\includegraphics[width=\linewidth]{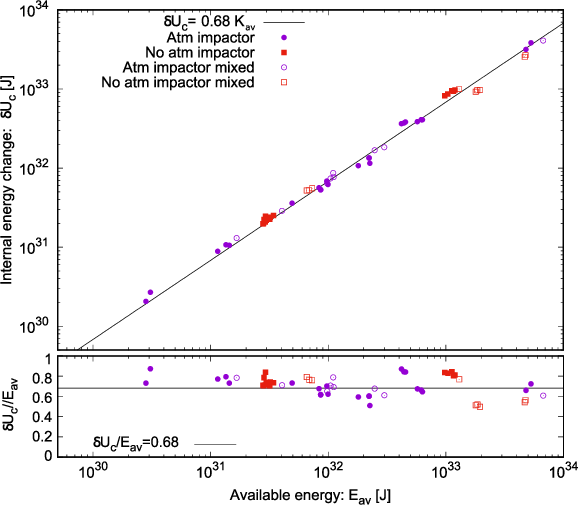}
\caption{The relationship between the available energy calculated by Eq.~\ref{Eav} and the internal energy change calculated by Eq.~\ref{deU1}.
The black line indicates the relation $\delta U_c=0.68~E_\mathrm{av}$.
The left bottom panel shows the fraction $\delta U_c$ to $E_\mathrm{av}$. Plot colors are the same as those in Figure~\ref{Kesc_hikaku}.
\label{PhidelU}}
\end{center}
\end{figure}

\begin{figure}[htbp]
\begin{center}
\includegraphics[width=\linewidth]{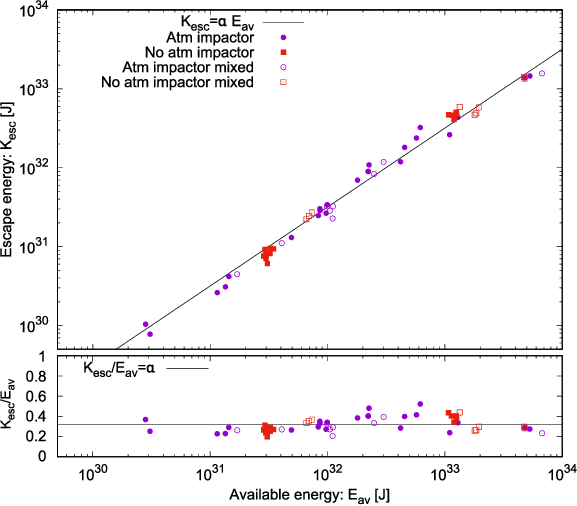}
\caption{The relationship between the available energy calculated by Eq.~\ref{Eav} and the kinetic energy of escaping particles. The black line indicates the relation $K_\mathrm{esc}=\alpha~E_\mathrm{av}$ where $\alpha=0.31$.
Plot colors are the same as those shown in Figure~\ref{Kesc_hikaku}.
\label{PhidelK}}
\end{center}
\end{figure}

Since the core undergoes an non-radial oscillation,
the remainder of the impact energy subtracted by the atmospheric expansion is expected to contribute to the heating of the core.
Figure~\ref{PhidelU} shows the ratio of $\delta U_c$ to $E_\mathrm{av}$. 
Our simulations indicate $\delta U_c = 0.68 E_\mathrm{av}$,
which means 68\% $E_\mathrm{av}$ transports to the core, and the rest 32\% of $E_\mathrm{av}$ is distributed in the atmosphere.
We find that Eq.~\ref{alpha-def} indicates $\alpha=0.32$. Next, we compare $K_\mathrm{esc}$ and $E_\mathrm{av}$. Figure~\ref{PhidelK} shows the ratio of $K_\mathrm{esc}$ to $E_\mathrm{av}$. Our simulation indicates that $K_\mathrm{esc} = 0.31 E_\mathrm{av}$, which means that 31\% of $E_\mathrm{av}$ is distributed to $K_\mathrm{esc}$. We find that Eq.~\ref{KescKinput} indicates $\alpha = 0.31$. The relative error of $\alpha$ between Eq.~\ref{alpha-def} and \ref{KescKinput} is 3\%, which is in good agreement. That is, we find that $E_\mathrm{av}$ is transformed to $\delta U_c$ by about 70\% and $K_\mathrm{esc}$ by 30\%.

Now, we can find that the kinetic energy of escaping particles is predicted by the energy conservation law if the planet undergoes a head-on collision. A scaling law for the atmosphere loss originates from the energy distribution between the released energy from the merged core, the atmosphere expansion, and the core heating.

\subsection{Empirical predictions for the atmosphere loss fraction}  \label{sec46}
\begin{figure}[htbp]
\begin{center}
\includegraphics[width=8cm]{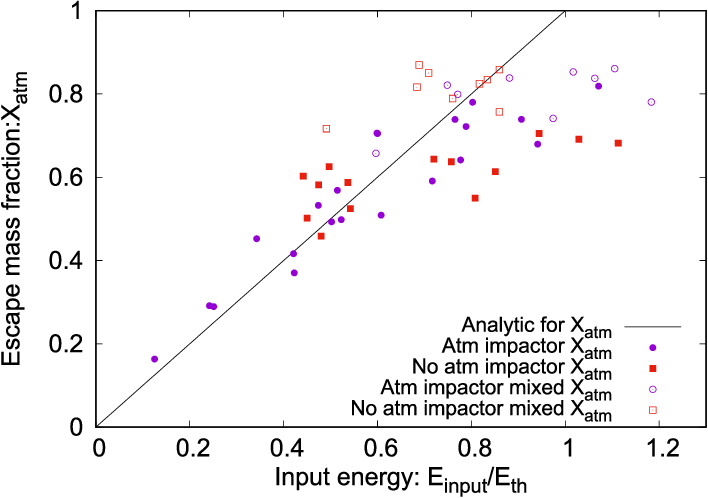}
\includegraphics[width=8cm]{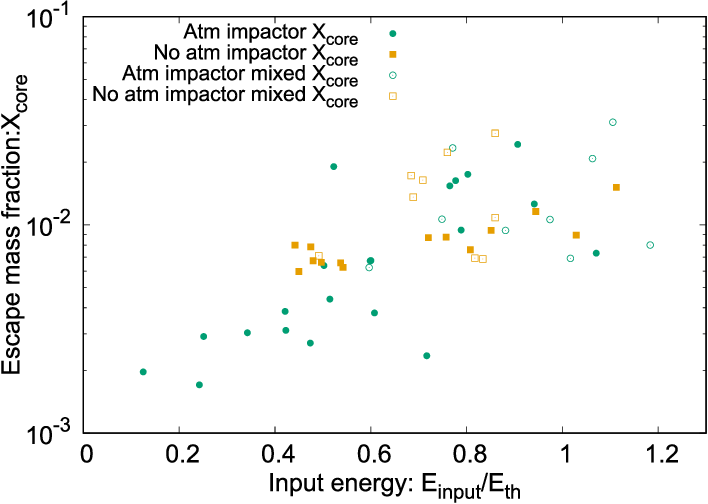}
\caption{
The relationship between the input energy and atmosphere loss fraction.
The escaping mass fraction for the atmosphere $X_\mathrm{atm}$ and core $X_\mathrm{core}$ are shown, which escape mass fractions are normalized by the total atmospheric mass $M_\mathrm{T,atm}+M_\mathrm{I,atm}$ and core mass $M_\mathrm{T,c}+M_\mathrm{I,c}$, respectively.
Circles are the impact simulations when the impactor has an atmosphere, while squares are the bare-rock impactor.
Filled marks reflect that the planet retains a layered structure after the impact, while open marks represent that the planet has a region where the atmosphere and core are mixed.
Purple and red symbols represent the atmosphere loss fraction $X_\mathrm{atm}$,
while green and orange symbols are the core loss mass fraction $X_\mathrm{core}$.
The black line is the analytic prediction for $X_\mathrm{atm}$ shown in Eq.~\ref{anal1-1}.
\label{KedW_Xatm}}
\end{center}
\end{figure}

We have derived two kinds of expression of $K_\mathrm{esc}$ by Eq.~\ref{Esc-bound} and Eq.~\ref{KescKinput}. Eliminating $K_\mathrm{esc}$ by Eq.~\ref{Esc-bound} and Eq.~\ref{KescKinput}; we can find the relationship for $M_\mathrm{esc}$ shown as
\begin{eqnarray}
\alpha E_\mathrm{av} = \beta \frac{GM_BM_\mathrm{esc}}{R_B}. \label{anal1} 
\end{eqnarray}
Since the escaping mass is dominated by the atmosphere, we focus on the atmosphere loss fraction shown as $M_\mathrm{esc} \approx M_\mathrm{esc,atm}$.
Since $E_\mathrm{av}$ includes the atmosphere loss fraction $X_\mathrm{atm}$,
we introduce the input energy $E_\mathrm{input}$ denoted by
\begin{equation}
E_\mathrm{input}  =   K_\mathrm{imp} - \delta \Phi_c + (1- \epsilon) (E_\mathrm{T,atm}+E_\mathrm{I,atm}). \label{Kinput-wH}
\end{equation}
Note that $E_\mathrm{input}$ is equivalent to $E_\mathrm{av}$ with $M_\mathrm{esc}=0$.
$E_\mathrm{input}$ is determined by the impact velocity and the internal structures of the target and impactor.
Thus, we find $X_\mathrm{atm}$ by Eq.~\ref{anal1} as
\begin{eqnarray}
X_\mathrm{atm} & = & \frac{E_\mathrm{input}}{E_\mathrm{th}}, \label{anal1-1} \\
E_\mathrm{th} & = & \frac{\beta}{\alpha} \frac{GM_B M_\mathrm{atm}}{R_B}- \epsilon (E_\mathrm{T,atm}+E_\mathrm{I,atm}),  \label{anal1-Ec}
\end{eqnarray}
where $E_\mathrm{th}$ is the threshold energy.
Eq.~\ref{anal1-1} is capped at 1 for the total atmosphere loss.
Using Eqs.~\ref{Esc-bound}, \ref{H-eng-def}, and \ref{KescKinput}, we put $\alpha = 0.31$, $\beta=0.42$ and $\epsilon=0.56$.
That is, we can find the relationship between $X_\mathrm{atm}$ and $E_\mathrm{input}$.
Figure~\ref{KedW_Xatm} shows the relationship between $X_\mathrm{atm}$ and $E_\mathrm{input}$.
Eq.~\ref{anal1-1} implies that all the atmosphere will escape if $E_\mathrm{input} > E_\mathrm{th}$ satisfied. 
However, our simulation results show that the atmosphere loss fraction approaches 1, but some results whose $E_\mathrm{input}$ exceeds $E_\mathrm{th}$ do not reach $X_\mathrm{atm}=1$. That is because the rock material begins to escape.

\begin{figure}[htbp]
\begin{center}
\includegraphics[width=\linewidth]{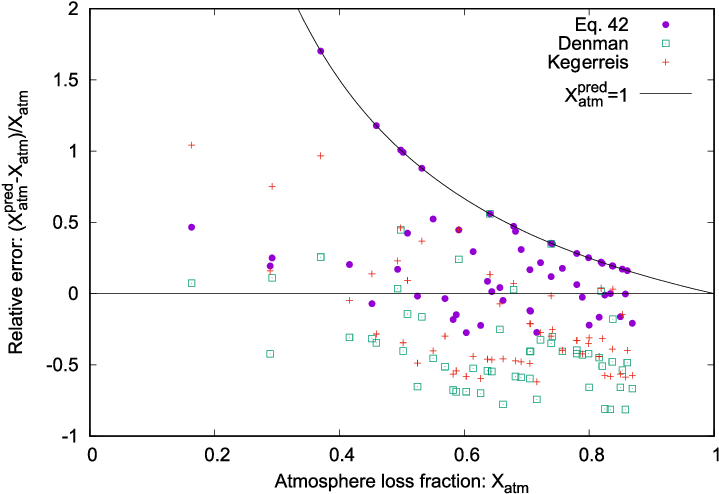}
\caption{
The relative error between the numerical results $X_\mathrm{atm}$ and the predictions $X_\mathrm{atm}^\mathrm{pred}$.
Purples marks are calculated by Eq.~\ref{MescMtot}.
Red marks are Eq.~1 in \citet{Kegerreis2020b}. Green marks are Eq.~21 in \citet{Denman2020}.
Black dashed line is the upper limit line for $X_\mathrm{atm}^\mathrm{pred}=1$.
\label{KedW_Mesc}}
\end{center}
\end{figure}

Conversely, when $E_\mathrm{input}$ is less than $E_\mathrm{th}$, the atmosphere loss fraction is proportional to the core energy change $\delta \Phi_c$ shown in Eq.~\ref{Phic}.
Combining Eq.~\ref{Mb-def}, \ref{Yb-def}, and \ref{anal1},
we find the predicted atmosphere loss $X_\mathrm{atm}^\mathrm{pred}=M_\mathrm{esc,atm}^\mathrm{pred}/M_\mathrm{atm}$ as 
\begin{eqnarray}
X_\mathrm{atm}^\mathrm{pred} &=& \frac{M_\mathrm{atm}}{2M_\mathrm{tot}} B \left[ 1-\sqrt{1-4\frac{E_\mathrm{input}}{B^2}\left( \frac{\beta}{\alpha}\frac{GM_\mathrm{tot}^2}{R_\mathrm{B}} \right)^{-1}}  \right] \label{MescMtot} \\
B &=& 1- \epsilon (E_\mathrm{T,atm}+E_\mathrm{I,atm}) \left(\frac{\beta}{\alpha}\frac{GM_\mathrm{tot}M_\mathrm{atm}}{R_\mathrm{B}} \right)^{-1} \label{MescMtotA}
\end{eqnarray}
where $M_\mathrm{tot}$ is the total mass which is described as $M_\mathrm{tot}=M_\mathrm{T}+M_\mathrm{I}$.
Figure~\ref{KedW_Mesc} shows the comparison between the numerical results $X_\mathrm{atm}$ and the analytical predictions  $X_\mathrm{atm}^\mathrm{pred}$ of the escaped atmospheric mass.

Our discussion assumes negligible escaping mass of the core. 
We consider the core loss fraction defined as Eq~\ref{Rloss}. 
Figure~\ref{KedW_Xatm} also shows the core loss fraction $X_\mathrm{core}$ for the core energy change.
We find that $X_\mathrm{core}$ increases with an increase in the $E_\mathrm{input}$.
In our simulation, the core loss fraction is less than $3\%$.
Thus, the core loss is much smaller than the atmosphere loss by one smaller order of magnitude.

We compare the numerical and the predicted atmosphere loss.
We consider the collision result for Run 28 C1+D1 shown in section \ref{exrock} as an example. Our numerical simulation showed that $X_\mathrm{atm} = 0.55$, while our prediction calculated by Eq.~\ref{MescMtot} gives $X_\mathrm{atm}^\mathrm{pred} = 0.84$.
We compare other predictions proposed by \citet{Denman2020} and \citet{Kegerreis2020b}. 
\citet{Denman2020} suggested that the atmosphere loss becomes dominant when the specific relative kinetic energy $(Q_R =\frac{1}{2}\mu v_\mathrm{imp}^2/M_\mathrm{tot})$ is below the transition value denoted as $Q_\mathrm{piv}$. In contrast, the core loss increases if the kinetic impact energy is larger than the transition. We set the impact velocity $v_\mathrm{imp} = v_\mathrm{esc,c}$ where $v_\mathrm{esc,c}$ is the mutual escape velocity from the core's surface denoted as $v_\mathrm{esc,c}=\sqrt{2G(M_\mathrm{T}+M_\mathrm{I})/(R_\mathrm{T,c}+R_\mathrm{I,c})}.$ Using Eq.~21 in \citet{Denman2020}, we find $X_\mathrm{atm}^\mathrm{D20}=0.45$. 
The meaning of $E_\mathrm{th}$ is equivalent to the transition energy $Q_\mathrm{piv}$ proposed by \citet{Denman2020}. Although the gravitationally bound strength of our target's planetary atmosphere is weaker than in the previous study, the atmosphere loss is consistent. 
\citet{Kegerreis2020b} studied the impact-induced atmosphere loss for planets whose atmosphere mass fractions is 1\% by mass.

Next, we compared all results with the prediction in Figure~\ref{KedW_Mesc}.
We checked the correlation coefficient between numerical results and predictions. The correlation coefficients for $X_\mathrm{atm}^\mathrm{D20}$, $X_\mathrm{atm}^\mathrm{K20}$, and Eq.~\ref{MescMtot} are 0.18, 0.28, and 0.50, respectively.
We find that when the atmosphere loss fraction increase, our study provide better prediction.
$X_\mathrm{atm}^\mathrm{D20}$ and $X_\mathrm{atm}^\mathrm{K20}$ typically underestimate the atmosphere loss while Eq.~\ref{MescMtot} typically overestimate.
This difference comes from the planetary structure and the thickness of the atmosphere.
The target we consider is less gravitationally bound by the core, while the Denman et al.'s target is a smaller radius than our target and \citet{Kegerreis2020b} considered thinner atmospheres.
Thus, the target loses the atmosphere more efficiently than in \citet{Denman2020}.

We describe how to use our formula to estimate the atmospheric loss from other head-on collisions in general.
Calculations of the target's and impactor's internal structures are necessary for our formula. Moreover, our formula also requires some iterative process for the trial atmospheric escape mass $M_\mathrm{esc}^\mathrm{trial}$.
We calculate $X_\mathrm{atm}^\mathrm{pred}$ using Eq.~\ref{MescMtot} and \ref{MescMtotA}, where $R_B$ included in the equations is calculated Eq.~\ref{Rb-def} by $M_B=M_T+M_I$.
We find escaping mass $M_\mathrm{esc,atm}^\mathrm{trial} = X_\mathrm{atm}^\mathrm{pred}\cdot (M_T+M_I)$ because $M_B$ is assumed to be $M_T+M_I$ for the first step of the iterative process.
Using $M_\mathrm{esc,atm}^\mathrm{trial}$, we find the $M_B = M_T+M_I - M_\mathrm{esc,atm}^\mathrm{trial}$.
Then we calculate the new $X_\mathrm{atm}^\mathrm{pred}$ using Eq.~\ref{MescMtot} and \ref{MescMtotA} with $R_B$ by $M_B=M_T +M_I-M_\mathrm{esc,atm}^\mathrm{trial}$. Thus, we find the new escaping mass $M_\mathrm{esc,atm}^\mathrm{trial}$. Finally, we check the consistency between old and new $M_\mathrm{esc,atm}^\mathrm{trial}$. Below we summarize the procedure of the calculation.
\begin{enumerate}
\item Estimate the interior structures for the target and the impactor to determine the internal energy and gravitational energies by Eq.~\ref{Uc-def}, \ref{Uatm-def}, \ref{Phic-def}, and \ref{Phiatm-def}. 
$\delta \Phi_c$ is calculated by Eq.~\ref{Phic}, which is determined that the bare-rock core is completely merged.
\item Set the trial atmospheric escape mass $M_\mathrm{esc,atm}^\mathrm{trial}$ and calculate $R_\mathrm{B}$ by using Eq.~\ref{Rb-def}.
\item Calculate the atmospheric escape mass by Eq.~\ref{anal1-1} and \ref{anal1-Ec}.
\item Check the convergence of the trial atmospheric escape mass $M_\mathrm{esc,atm}^\mathrm{trial}$ and $M_\mathrm{esc,atm}^\mathrm{pred}$.
\end{enumerate}
Note that the atmosphere loss fraction $X_\mathrm{atm} = 1$ when $E_\mathrm{input} > E_\mathrm{th}$.

\section{Discussion} \label{sec:discussion}

\subsection{Implications for the origin of planets with massive atmospheres}

In the protoplanetary disk, protoplanets undergo collision-coalescence processes, and their solid cores grow.
During planetary accretion, larger planetesimals are supposed to grow faster than the smaller ones, which is the runaway growth of the largest planetesimal because of the gravitational focusing and dynamical friction from smaller planetesimals \citep[e.g.,][]{Greenberg1978,Wetherill1989,Kokubo1996}.
Subsequently, the largest planetesimal becomes a protoplanet.
After the runaway growth, the protoplanet repels its neighbors to a distance wider than a 5~Hill radius, which means that the growth mode shifts toward oligarchic growth.
The final mass is estimated as the isolation mass in the oligarchic growth stage. 
That is, Mars-size planetary impact is an essential path for forming Earth-size planets
\citep[e.g.,][]{Chambers1998, Ogihara2018, Kokubo2002, Goldreich2004}.
According to the observation of exoplanets, there are many kinds of exoplanets with large planetary radii, which suggests they have a massive atmosphere \citep{Rogers2015,Fulton2018,Rogers2021}.
The growing cores obtain primordial atmospheres from the gas in the protoplanetary disk.
Accretion of the atmosphere onto each protoplanet may continue until the protoplanetary disk disappears.
As a result, the protoplanets are possibly expected to have a primordial atmosphere.
After the disk dissipation, protoplanets undergo coalescence.
Their core masses grow efficiently while the atmospheres are lost in this process.

Since oblique impacts occur more frequently and remove significantly less atmosphere than head-on impacts, 
the oblique impact will not contribute to planetary growth in general. \citep{Kegerreis2020b,Denman2022}.
Although this study deals with head-on impacts, it is expected to be a good approximation for minor angle collisions. In the following discussion, we consider the possibility of forming planets with atmospheres of 10~\% or more, assuming near-head-on collisions dominates in the formation of such planets.

The probability of high-velocity collisions, which are higher than $2v_\mathrm{esc}$, is about 10\% according to dynamical simulations of protoplanets at the late-stage planetary accretion. \citep[e.g.,][]{Raymond2009, Stewart2012, Matsumoto2021, Ogihara2021}. 
Our simulations demonstrate that
Our simulations demonstrate that in the case of the planet with 10~\% of the atmosphere only an atmosphere of a few percent in mass remains even after a single impact event.
Exoplanets with large radii, such as super-puff planets, have been reported among the currently observed exoplanets. Such exoplanets are expected to have a large amount of atmosphere.
If such a planet has an atmospheric mass larger than 10~\%, the planet should have avoided the significant atmosphere loss by a high-energy, head-on collision.

Another possibility is the accretion event with many small-sized bodies, such as the pebble accretion \citep[e.g.,][]{Lambrechts2012,Kurokawa2018}.
When the planet is formed without a giant impact or with relatively low-energy $(v_\mathrm{imp}\lesssim 2 v_\mathrm{esc})$ and/or oblique impacts, the formed planet may avoid significant impact-induced atmosphere loss.
Note, however, that
the pebble accretion process has problems of the low accretion efficiency and the long core growth timescale \citep{Ormel2010, Okamura2021}.

\subsection{The growth of the rocky core: comparing to the bare-rock accretion}
In this subsection, we focus on the core mass after the impact. 
In all of our simulations, the resulting gravitationally bound mass is larger than the original mass of the target. The target's and impactor's cores merged with negligible core loss, while the atmosphere escapes significantly.
\citet{Agnor2004} calculated the impact simulation for rock-composed protoplanets without the atmosphere and indicated that the protoplanets merging is less efficient if the impact velocity is higher than 1.5~escape velocity in the case of a head-on collision.
Even though our results include a head-on collision impact with the relative velocity higher than 1.5~escape velocity, 
the core can merge into a single object with negligible core loss when the protoplanet has an atmosphere.
Thus, we have an important conclusion that the atmosphere facilitates the core mass growth. 
In the case of oblique impact events, previous studies suggested that a so-called hit-and-run collision does not contribute to the core mass growth \citep{Agnor2004,Genda2012,Leinhardt2012}.
However, the planetary atmosphere can help merge cores.
For the oblique impact of ice-giants, 
the core would merge completely despite the large impact angle \citep[e.g.][]{Kegerreis2018,Chau2018,Kurosaki2019}.
That is because the planetary atmosphere increases planetary radius.
Although our study considered only head-on impact, the effects of atmosphere loss and angular momentum transport in oblique impact are important issues to be addressed in the future.

\subsection{Post-impact interior structure with the mix of the atmosphere and core} \label{miximpact}

\begin{figure}[htbp]
\begin{center}
\includegraphics[width=\linewidth]{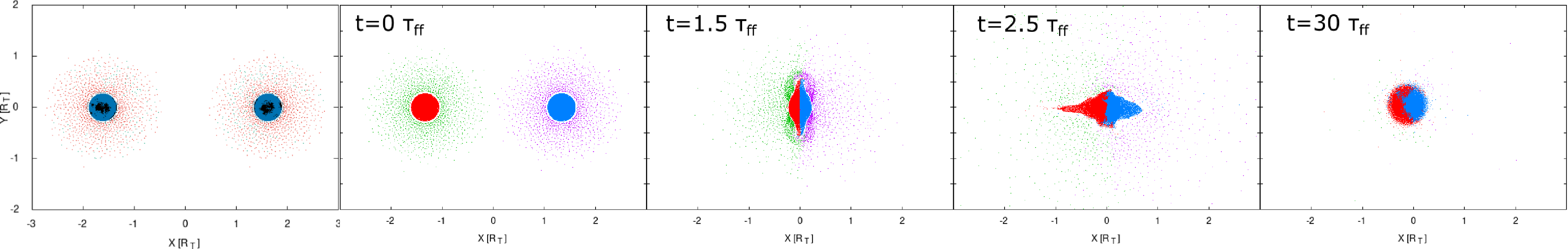}
\caption{
Snapshots of impact simulations. 
The top left-most panel is color-coded diagrams of lost and bound particles.
Red indicates lost atmospheric particles. Green indicates bound atmospheric particles. Black indicates escaping rock particles, and blue indicates bound rock particles.
For the target mass of $1~M_\oplus$ with 10\% of the atmosphere by mass and the impactor mass of $1~M_\oplus$ with 10\% of the atmosphere (Mix~34 A3+A3), simulation results of central cross sections are shown in four panels.
Those snapshots shows the time when $t=0\times\tau_\mathrm{ff}$ (Panel 1), $t=1.5\times\tau_\mathrm{ff}$ (Panel 1), $t=2.5\times\tau_\mathrm{ff}$ (Panel 2), and $t=30\times\tau_\mathrm{ff}$ (Panel 4) from left to right, respectively.
Run~31 simulation is start from $t=-2.5\times\tau_\mathrm{ff}$, where $\tau_\mathrm{ff}=2.2~$hours.
The colors are the same as in Figure~\ref{12figexample}.
 \label{mixfigexample}}
\end{center}
\end{figure}

\begin{figure}[htbp]
\begin{center}
\includegraphics[width=7cm]{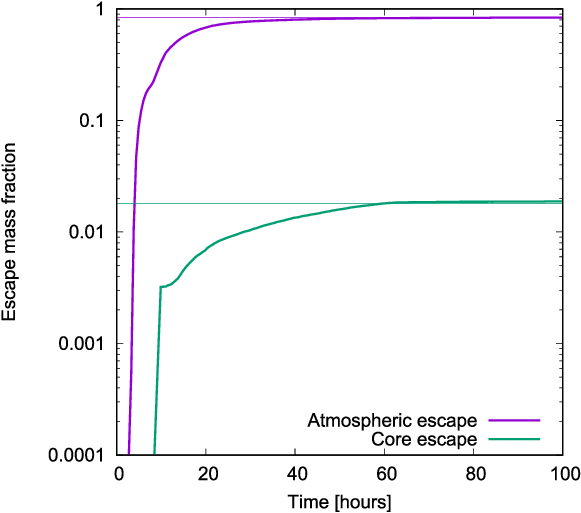}
\includegraphics[width=8cm]{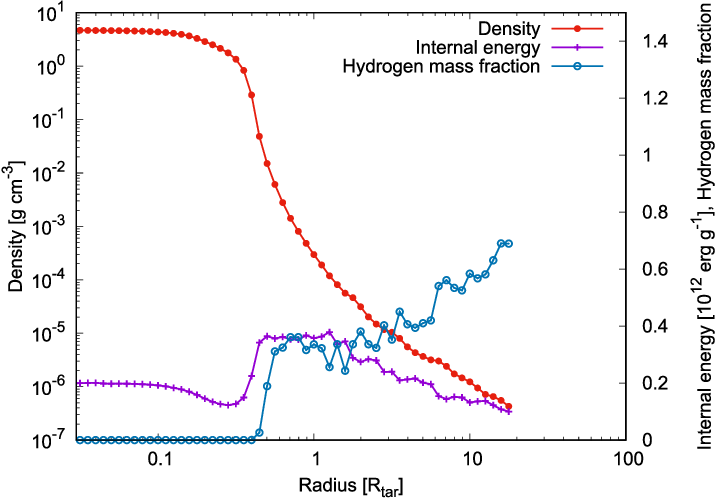}
\caption{Left panel: Time evolution of the escaping mass fraction after the impact event.
The target mass is $1~M_\oplus$ with 10\% of the atmosphere by mass. The impactor is the same as the target. 
Colored lines are the same as those in the Figure~\ref{12lossexample}.
Right panel: The interior structure at $t = 50~\tau_\mathrm{ff}$. The target mass is $1~M_\oplus$ with 10\% of the atmosphere by mass while the impactor is the same as the target.  
The bound body radius is normalized by the target radius.
Colored lines are the same as those in Figure~\ref{12interior}.
\label{mixinterior}}
\end{center}
\end{figure}

In the relaxation after the giant impact, the rocky material is expected to fall onto the center of the resultant body
because SPH rock particles tend to have larger densities suggests that they are likely to settle down.
On the other hand, some of our simulation results show that the rock component  evaporates and mixes with the atmospheric component after the impact. To explain this, we choose Mix 34 A3+A3 in Table~\ref{impact_results}, i.e., the $1~M_\oplus$ target and impactor with 10\% of the atmosphere colliding at 1.0 escape velocity. Figure~\ref{mixfigexample} shows the simulation snapshots. The atmosphere escapes as a result of the core oscillating after it merges. 
The left panel of Figure~\ref{mixinterior} shows the time evolution for the atmosphere and core loss fractions, while the right panel of figure~\ref{mixinterior} shows the interior structure after the impact.
The right panel of figure~\ref{mixinterior} shows that the atmosphere and core are mixed where $y_\mathrm{B} (r=R_\mathrm{T}+R_\mathrm{I}) = 0.33$, unlike the results in figures~\ref{12interior} and \ref{34interior}.
The mixed profile is stable from 10 to 50 $\tau_\mathrm{ff}$. The loss fractions for the atmosphere and core in the case are 84\% and 1.9 \%, respectively.
The names of the runs starting with "Mix"  in Table~\ref{impact_results} reflect the resulting structure that shows this mixed structure. Those results suggest that the dredged-up rock vapor and the atmospheric component are stable in a mixed state for our simulation time scale. Our calculations do not consider the radiative cooling and cannot discuss the long-term stability of the planetary structure. However, the detailed planetary structure model that considers a planet consisting of a hydrogen atmosphere atop a magma ocean implies that the core will be hot enough to evaporate due to the blanketing effect of the hydrogen atmosphere covering the core \citep[e.g.,][]{Abe1997}. Therefore, there is a possibility of mixed atmospheric and rock vapor after the impact.

Figure~\ref{KedW_Xatm} also shows the relationship between $E_\mathrm{input}$ and $K_\mathrm{esc}$ for post-mixed simulation results.
We find that the energy budget between $E_\mathrm{input}$ and $K_\mathrm{esc}$ seems similar to the case without the mixed structure when the core and atmosphere are mixed.
As a result, the kinetic energy of the escaping particles can be reproduced by Eq.~\ref{Eav}. Thus, we can predict the atmosphere loss after the impact based on figure~\ref{KedW_Xatm}. When the collision injects significant energy into the merged core, the remainder energy ejects the atmosphere and oscillates the merged core. After the collision, the impact energy dredges the core vapor material into the atmosphere, causing the mixture of rock vapor and atmosphere. Thus, the rock vapor tends to mix more easily with the atmosphere when the impact energy is higher.

The impact-induced mixed of core material vapor might affect the atmospheric structure of exoplanets. The core material vapor condenses as the planetary atmosphere cools, resulting in a highly metallic atmosphere, as observed for certain super-Earths, such as GJ1214b \citep[e.g.][]{Morley2015,Ohno2020} and GJ436b \citep[e.g.,][]{Lothringer2018,Morley2017}. The rock vapor facilitates the reaction between the atmosphere and core material. Since the impact-induced rock vapor can react with hydrogen, the planetary atmosphere may become water-dominated \citep{Herbort2020,Schlichting2021,Itcovitz2022}. Condensation of the core material in the atmosphere may change the planet's thermal evolution because the latent heat raises the atmospheric temperature that determines the planetary luminosity \citep{Kurosaki2017}. Moreover, rapid condensation of rock components and subsequent sedimentation can remove the core material from the planetary atmosphere. That is, the giant impact significantly impacts the planetary atmospheric property, which will significantly impact the thermal evolution and the observation of the atmosphere.

\section{Conclusion} \label{sec:conclusion}
In this article, we have performed head-on giant impact simulations via smoothed particle hydrodynamic simulations to study the impact-induced atmosphere loss.
We chose a target with a mass of 0.1--10 Earth masses with 10\%, 20\%, and 30\% of the atmosphere while the impactor mass is 0.1--10~Earth masses with 0\%, 10\%, 20\%, and 30\% of the atmosphere.
We also investigated the post-impact internal structure and categorized results into  layered or mixed structures.
We analyzed the energy budget before and after the impact and investigated the relationship between the sum of kinetic and released energies from the merged core, the kinetic energy of escaping particles, and the atmosphere loss fraction.
 We summarize our results as follows:
\begin{itemize}
\item If the impactor mass is comparable to the target mass, almost all the core material from both objects merge after the collision and release a significant amount of self-gravitational energy that cause atmosphere loss from the target.
\item The kinetic energy of the escaping particles is nearly proportional to the sum of the kinetic energy prior to a collision in the center of the mass frame and released self-gravitational energy from the merged core.
The proportionality constant is mainly determined by the internal energy change of the core.
\item About 30\% of the sum of the impact energy and the released energy from the merged core
contributes to the atmosphere loss,
while 70\% of the energy is absorbed by the atmospheric expansion and the core heating.
\item When the impactor and target masses are comparable $(M_I \gtrsim 0.3 M_T)$ in the case of head-on collision, the atmosphere loss fraction is larger than 50\%.
\item The escaping materials are mostly the atmospheric gases.
Even during significant atmosphere loss, the core loses only 1\% of the total mass. 
Thus, the giant impact event studied in this article enhances the core growth at the expense of the atmosphere loss.
\item When the atmosphere loss fraction exceeds 70\%,
we observed in some numerical simulations that the core material vapor was mixed with the planetary atmosphere.
Even under those cases, the relationship between the remainder energy represented by the sum of the kinetic impact energy and the released energy from the merged core, the kinetic energy of escaping particles, and the atmosphere loss fraction appear to be similar to the situation in which the core material and atmosphere are not mixed.
\end{itemize}
In this study, we find that the larger the impactor mass, the more the atmosphere losses during the coalescence growth of the cores.
Our study shows that an impact event with an impactor mass much smaller than a target mass $(M_\mathrm{I} < 0.3 M_\mathrm{T})$ can retain the planetary atmosphere $(M_\mathrm{atm} > 0.1 M_\mathrm{B})$ even if the planet has undergone the head-on impact,
while the impact event with the comparable impactor mass results in a significant loss of the atmosphere.
Thus, the impact-induced atmosphere loss will be a key process in the study of the origin of the planet with a thin atmosphere that is grown by giant impacts after the dispersal of the protoplanetary disk \citep[e.g.,][]{Chambers1998, Ogihara2018, Kokubo2002, Goldreich2004}.
Detailed analyses of the oblique impact cases are expected to help our understanding of the relationship between the growth of planets and the atmosphere during planet formation.

\begin{acknowledgments}
The authors thank anonymous referees for various constructive comments.
The authors thank H. Kobayashi for the fruitful discussion and suggestions.
The authors also thank K. Kurosawa for suggestions regarding the equation of states. 
Our simulation code utilized FDPS.
Numerical computations were carried out on Cray XC50 at the Center for Computational Astrophysics, National Astronomical Observatory of Japan.
This work is supported by JSPS Grants-in-Aids for Scientific Research No.  20J01258, 21H00039 (KK), 16H02160, 18H05436, and 18H05437 (SI).
\end{acknowledgments}

\startlongtable
\begin{deluxetable*}{cccccccccccccccc}
\tablecaption{The list of the numerical result. 
We show the models named T+I where T is the target name, and I is the impactor name shown in Table~\ref{planetary_model}.
We show
the impact velocity normalized by the escape velocity $v_\mathrm{imp} [v_\mathrm{esc}]$,
the released energy from the merged core $\delta\Phi_c~[10^{38} \mathrm{erg}]$,
the remainder energy $E_\mathrm{input}~[10^{38} \mathrm{erg}]$,
the change of the internal energy of the core $\delta U_c~[10^{38} \mathrm{erg}]$,
the change of the atmospheric energy $\delta E_\mathrm{atm}~[10^{38}~\mathrm{erg}]$,
the threshold energy $ E_\mathrm{th}~[10^{38}\mathrm{erg}]$,
the kinetic energy of escaping particles $K_\mathrm{esc}~[10^{38} \mathrm{erg}]$,
the atmosphere loss fraction $X_\mathrm{atm}$,
the core loss fraction $X_\mathrm{core}$,
the gravitationally bound body's mass $M_B~[M_\oplus]$,
the radius after the impact $R_B~[R_\oplus]$,
the gravitationally bound body's atmospheric mass fraction $Y_B$,
the hydrogen mass fraction at the surface $y_B(R_s)$ where $R_s = R_\mathrm{T}+R_\mathrm{I}$,
and the predicted atmosphere loss $X_\mathrm{atm}^\mathrm{pred}$.
The Run numbers 1--33 reflect that the bound body keeps a layered structure as shown in figure~\ref{12interior}, and \ref{34interior} while the Run numbers 34--55, named Mix, are ones wherein the bound body's interior is mixed with the atmosphere and core as shown in figure~\ref{mixinterior} as an example.
\label{impact_results}}
\tablewidth{700pt}
\tabletypesize{\scriptsize}

\tablehead{
\colhead{Run} & \colhead{T+I} & \colhead{$v_\mathrm{imp}$} & \colhead{$\delta\Phi_c$ } & \colhead{$E_\mathrm{input}$ }  & \colhead{$\delta U_c$} &\colhead{$\delta E_\mathrm{atm}$} &  \colhead{$E_\mathrm{th}$} &  \colhead{$K_\mathrm{esc}$}  &  \colhead{\textbf{X}$_\textbf{atm}$} &  \colhead{$X_\mathrm{core}$} & \colhead{$M_B$}  &  \colhead{$R_B$}  & \colhead{$Y_B$} & \colhead{$y_B(R_s)$} & \colhead{$X_\mathrm{atm}^\mathrm{pred}$}
}

\startdata
Run   1 & A3+A1    &     1.67     &     1.44     &     1.38     &    -1.07    & -0.1594     &    2.679     &    0.309     &         \textbf{0.57} &       {0.0044} &     1.03     &     2.19     &        0.046 &  0.96 &  0.55     \\
Run   2 & A3+A1    &     2.31     &    0.980     &     1.75     &    -1.30    & -0.1780     &    2.676     &    0.447     &         \textbf{0.66} &       {0.0062} &     1.02     &     1.97     &        0.037 &  0.96 &  0.69     \\
Run   3 & B2+B2    &     1.26     &     9.91     &     9.95     &    -6.84    & -0.7431     &    10.57     &     2.65     &         \textbf{0.68} &       {0.011} &     1.71     &     3.34     &        0.075 &  0.90 &  0.83     \\
Run   4 & B3+B2    &     1.65     &     23.5     &     23.1     &    -13.4    &  -2.819     &    38.52     &     8.98     &         \textbf{0.71} &       {0.0067} &     3.41     &     3.65     &        0.069 &  0.96 &  0.62     \\
Run   5 & B3+B2    &     1.65     &     23.5     &     23.1     &    -13.4    &  -2.817     &    38.45     &     8.97     &         \textbf{0.70} &       {0.0067} &     3.41     &     3.66     &        0.069 &  0.94 &  0.62     \\
Run   6 & B4+B2    &     1.65     &     52.2     &     40.4     &    -38.2    &  -31.43     &    160.8     &     18.1     &         \textbf{0.29} &       0.0019 &     10.3     &     7.87     &        0.151 &  0.98 &  0.35     \\
Run   7 & B4+B2    &     2.30     &     27.4     &     62.2     &    -40.9    &  -34.86     &    181.5     &     32.3     &         \textbf{0.45} &       0.0030 &     9.98     &     6.63     &        0.121 &  0.99 &  0.42     \\
Run   8 & B2+B1    &     1.67     &     1.31     &     1.14     &   -0.886    & -0.2700     &    2.701     &    0.261     &         \textbf{0.42} &       0.0038 &     1.00     &     4.44     &        0.127 &  1.00 &  0.50     \\
Run   9 & B2+B1    &     2.31     &    0.958     &     1.46     &    -1.05    & -0.3099     &    2.916     &    0.419     &         \textbf{0.49} &       0.0064 &    0.986     &     4.00     &        0.113 &  1.00 &  0.58     \\
Run  10 & C1+C1    &     1.28     &     8.78     &     8.98     &    -5.30    &  -1.048     &    11.38     &     2.86     &         \textbf{0.72} &       0.0094 &     1.55     &     4.28     &        0.107 &  0.98 &  0.88     \\
Run  11 & C1+C1    &     1.26     &     8.15     &     8.98     &    -5.33    &  -1.086     &    11.74     &     3.02     &        \textbf{0.74} &       0.0154 &     1.53     &     4.10     &        0.102 &  0.95 &  0.83     \\
Run  12 & C2+C1    &     1.65     &     19.6     &     18.4     &    -10.7    &  -3.647     &    30.21     &     6.94     &        \textbf{0.51} &       0.0038 &     3.38     &     7.23     &        0.174 &  1.00 &  0.73     \\
Run  13 & C2+C1    &     2.30     &     12.0     &     23.3     &    -11.5    &  -3.888     &    32.50     &     10.9     &        \textbf{0.59} &       0.0024 &     3.28     &     6.49     &        0.149 &  0.98 &  0.86     \\
Run  14 & C3+C1    &     1.65     &     48.2     &     27.1     &    -36.5    &  -42.80     &    216.4     &     11.9     &        \textbf{0.16} &       0.0020 &     10.4     &     9.09     &        0.264 &  0.99 &  0.24     \\
Run  15 & C3+C1    &     2.30     &     29.8     &     49.3     &    -38.7    &  -48.97     &    203.6     &     23.8     &        \textbf{0.29} &       0.0017 &     10.0     &     9.36     &        0.233 &  0.99 &  0.37     \\
Run  16 & A5+A3    &     1.66     &     43.7     &     111.     &    -37.1    &  -4.735     &    177.2     &     26.2     &        \textbf{0.66} &       0.0019 &     10.3     &     3.16     &        0.036 &  0.97 &  0.63     \\
Run  17 & A4+A3    &     2.30     &     11.7     &     30.6     &    -18.3    &  -1.210     &    26.22     &     11.8     &        \textbf{0.84} &       0.0094 &     3.63     &     2.10     &        0.018 &  0.97 &   1.0     \\
Run  18 & A1+A1    &     1.64     &    0.304     &    0.310     &   -0.269    & -0.0064     &   0.3424     &    0.077     &        \textbf{0.74} &       0.0243 &    0.181     &     1.05     &        0.029 &  1.00 &   1.0     \\
Run  19 & B4+B4    &     1.64     &     534.     &     536.     &    -383.    &  -82.30     &    1131.     &     145.     &         \textbf{0.53} &       0.0027 &     17.8     &     6.41     &        0.105 &  0.92 &   1.0     \\
Run  20 & B1+B1    &     1.64     &    0.281     &    0.284     &   -0.206    & -0.0021     &   0.3658     &    0.104     &        \textbf{0.64} &       0.0163 &    0.172     &     1.88     &        0.084 &  1.00 &   1.0     \\
Run  21 & C1+C1    &     1.64     &     8.58     &     8.42     &    -5.65    & -0.7773     &    16.10     &     2.47     &        \textbf{0.50} &       0.0191 &     1.67     &     6.41     &        0.180 &  0.95 &   1.0     \\
Run  22 & C3+C3    &     1.64     &     492.     &     472.     &    -317.    &  -121.7     &    1117.     &     137.     &        \textbf{0.37} &       0.0031 &     17.7     &     9.91     &        0.213 &  0.95 &   1.0     \\
Run  23 & B2+D1    &     1.00     &     3.36     &     2.95     &    -1.98    & -0.3760     &    4.906     &    0.762     &        \textbf{0.64} &       0.0087 &     1.07     &     2.74     &        0.066 &  0.98 &  0.65     \\
Run  24 & B2+D1    &     1.09     &     3.23     &     3.07     &    -2.08    & -0.3737     &    4.856     &    0.808     &        \textbf{0.64} &       0.0087 &     1.07     &     2.77     &        0.068 &  0.97 &  0.69     \\
Run  25 & B2+D1    &     1.41     &     3.00     &     3.33     &    -2.25    & -0.3567     &    4.688     &    0.945     &       \textbf{0.61} &       0.0094 &     1.08     &     2.89     &        0.072 &  0.96 &  0.79     \\
Run  26 & B4+D3    &     1.00     &     153.     &     122.     &    -94.9    &  -38.28     &    269.4     &     40.8     &        \textbf{0.52} &       0.0063 &     11.0     &     5.22     &        0.086 &  0.95 &  0.52     \\
Run  27 & B4+D3    &     1.11     &     142.     &     131.     &    -97.0    &  -37.65     &    290.6     &     47.4     &        \textbf{0.59} &       0.0066 &     10.9     &     4.76     &        0.076 &  0.95 &  0.50     \\
Run  28 & C1+D1    &     1.00     &     3.46     &     3.00     &    -1.98    & -0.4299     &    4.427     &    0.919     &       \textbf{0.55} &       0.0042 &     1.04     &     4.53     &        0.130 &  1.00 &  0.84     \\
Run  29 & C1+D1    &     1.09     &     3.26     &     4.87     &    -2.24    & -0.3858     &    4.198     &    0.711     &        \textbf{0.50} &       0.0060 &     1.05     &     4.88     &        0.142 &  0.99 &   1.0     \\
Run  30 & C1+D1    &     1.40     &     2.90     &     4.94     &    -2.45    & -0.3858     &    4.027     &    0.612     &       \textbf{0.46} &       0.0067 &     1.07     &     5.16     &        0.152 &  0.98 &   1.0     \\
Run  31 & C3+D3    &     1.00     &     154.     &     116.     &    -82.0    &  -56.38     &    312.5     &     46.9     &        \textbf{0.60} &       0.0065 &     10.2     &     6.47     &        0.116 &  0.97 &  0.44     \\
Run  32 & C3+D3    &     1.10     &     143.     &     122.     &    -86.4    &  -56.86     &    304.2     &     46.0     &        \textbf{0.58} &       0.0078 &     10.3     &     6.71     &        0.122 &  0.97 &  0.48     \\
Run  33 & C3+D3    &     1.42     &     124.     &     135.     &    -93.6    &  -56.63     &    322.9     &     50.6     &        \textbf{0.63} &       0.0066 &     10.2     &     6.21     &        0.110 &  0.97 &  0.49     \\
Mix  34 & A3+A3    &     1.00     &     11.6     &     10.7     &    -7.39    & -0.4180     &    5.144     &     2.85     &        \textbf{0.84} &       0.019 &     1.80     &     1.80     &        0.018 &  0.33 &   1.0     \\
Mix  35 & A3+A3    &     1.28     &     10.4     &     11.2     &    -8.68    & -0.3988     &    4.863     &     2.27     &        \textbf{0.78} &       0.0080 &     1.83     &     1.95     &        0.024 &  0.63 &   1.0     \\
Mix  36 & A3+A3    &     1.26     &     9.57     &     11.3     &    -7.69    & -0.4240     &    5.249     &     3.24     &        \textbf{0.86} &       0.0311 &     1.77     &     1.74     &        0.016 &  0.71 &   1.0     \\
Mix  37 & A4+A3    &     1.65     &     24.8     &     25.4     &    -16.8    &  -1.225     &    25.64     &     8.31     &        \textbf{0.82} &       0.0106 &     3.63     &     2.15     &        0.020 &  0.41 &  0.99     \\
Mix  38 & A3+A2    &     2.30     &     2.66     &     5.02     &    -3.61    & -0.2246     &    4.687     &     1.30     &        \textbf{0.82} &       0.0073 &     1.19     &     1.66     &        0.020 &  0.82 &   1.0     \\
Mix  39 & A5+A3    &     2.31     &     13.7     &     130.     &    -40.8    &  -3.529     &    206.6     &     43.5     &        \textbf{0.80} &       0.0047 &     10.1     &     2.66     &        0.022 &  0.88 &  0.62     \\
Mix  40 & A3+A2    &     1.65     &     4.12     &     4.15     &    -2.87    & -0.2102     &    3.307     &     1.11     &        \textbf{0.74} &       0.0106 &     1.19     &     1.85     &        0.028 &  0.81 &   1.0     \\
Mix  41 & B2+B2    &     1.28     &     9.81     &     10.3     &    -6.20    & -0.8019     &    12.78     &     3.40     &        \textbf{0.78} &       0.0175 &     1.66     &     2.66     &        0.053 &  0.87 &   1.0     \\
Mix  42 & B2+B2    &     1.64     &     9.02     &     10.2     &    -6.39    & -0.8147     &    6.867     &     3.25     &        \textbf{0.80} &       0.0234 &     1.64     &     2.54     &        0.049 &  0.54 &   1.0     \\
Mix  43 & A5+A5    &     1.64     &     512.     &     542.     &    -409.    &  -6.982     &    664.7     &     157.     &        \textbf{0.85} &       0.0069 &     18.2     &     2.70     &        0.016 &  0.40 &   1.0     \\
Mix  44 & B4+D3    &     1.43     &     117.     &     143.     &    -99.7    &  -36.17     &    291.8     &     58.8     &        \textbf{0.72} &       0.0071 &     10.6     &     3.82     &        0.053 &  0.83 &  0.52     \\
Mix  45 & A3+D1    &     1.00     &     3.42     &     3.06     &    -2.19    & -0.1937     &    3.895     &    0.747     &       \textbf{0.70} &       0.0116 &     1.13     &     1.78     &        0.026 &  0.86 &  0.82     \\
Mix  46 & A3+D1    &     1.10     &     3.35     &     3.30     &    -2.39    & -0.1876     &    3.855     &    0.820     &       \textbf{0.69} &       0.0089 &     1.13     &     1.80     &        0.027 &  0.75 &  0.91     \\
Mix  47 & A3+D1    &     1.42     &     2.95     &     3.51     &    -2.52    & -0.1888     &    3.795     &    0.936     &       \textbf{0.68} &       0.0151 &     1.13     &     1.82     &        0.028 &  0.82 &  0.98     \\
Mix  48 & C1+D2    &     1.00     &     8.60     &     6.84     &    -5.20    & -0.5579     &    10.00     &     2.22     &        \textbf{0.82} &       0.0173 &     1.43     &     2.19     &        0.039 &  0.80 &  0.68     \\
Mix  49 & C1+D2    &     1.10     &     8.17     &     7.15     &    -5.26    & -0.5394     &    9.410     &     2.44     &        \textbf{0.79} &       0.0223 &     1.43     &     2.33     &        0.044 &  0.76 &  0.77     \\
Mix  50 & C1+D2    &     1.42     &     7.13     &     7.55     &    -5.55    & -0.5557     &    8.787     &     2.70     &        \textbf{0.76} &       0.0275 &     1.43     &     2.51     &        0.051 &  0.81 &  0.89     \\
Mix  51 & A5+D3    &     1.00     &     131.     &     181.     &    -92.3    &  -3.520     &    221.4     &     46.9     &        \textbf{0.82} &       0.0069 &     11.2     &     2.51     &        0.016 &  0.64 &  0.82     \\
Mix  52 & A5+D3    &     1.11     &     117.     &     187.     &    -96.4    &  -3.312     &    223.7     &     48.9     &        \textbf{0.83} &       0.0069 &     11.2     &     2.48     &        0.015 &  0.60 &  0.83     \\
Mix  53 & A5+D3    &     1.43     &     88.4     &     196.     &    -97.1    &  -3.323     &    228.3     &     58.0     &        \textbf{0.86} &       0.0109 &     11.1     &     2.41     &        0.013 &  0.55 &  0.86     \\
Mix  54 & C3+D4    &     1.00     &     406.     &     495.     &    -254.    &  -56.77     &    698.3     &     141.     &        \textbf{0.85} &       0.0164 &     14.2     &     3.14     &        0.032 &  0.70 &  0.71     \\
Mix  55 & C3+D4    &     1.12     &     356.     &     502.     &    -268.    &  -55.78     &    729.5     &     135.     &        \textbf{0.87} &       0.0136 &     14.2     &     3.00     &        0.028 &  0.55 &  0.69     \\
\enddata

\end{deluxetable*}

\appendix
\section{Distributions of escaping particles' velocity} \label{app-esc}
\begin{figure}[htbp]
\begin{center}
\includegraphics[width=\linewidth]{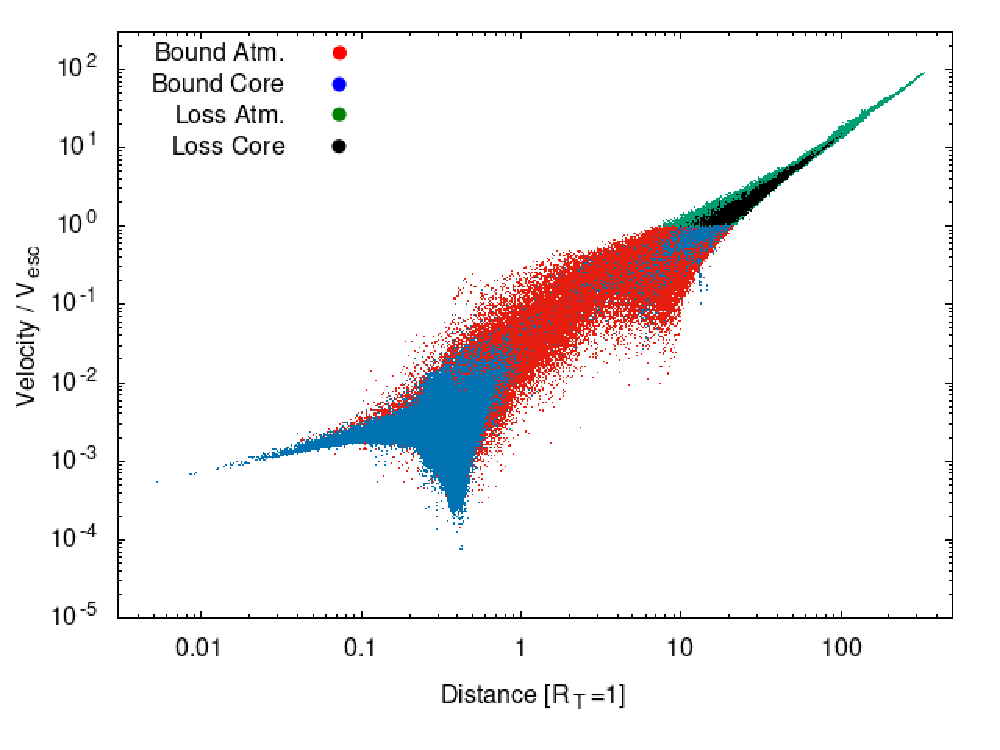}
\caption{
Upper panel: The velocity distribution for Run~31 after the impact. The velocity is normalized by Eq.~\ref{vesc-rad}. Red dots are the bound atmosphere particles. Blue dots are the bound core particles. Green particles are escaping atmosphere particles. Black particles are escaping core particles. \\
\label{vel_rad}}
\end{center}
\end{figure}

\begin{figure}[htbp]
\begin{center}
\includegraphics[width=14cm]{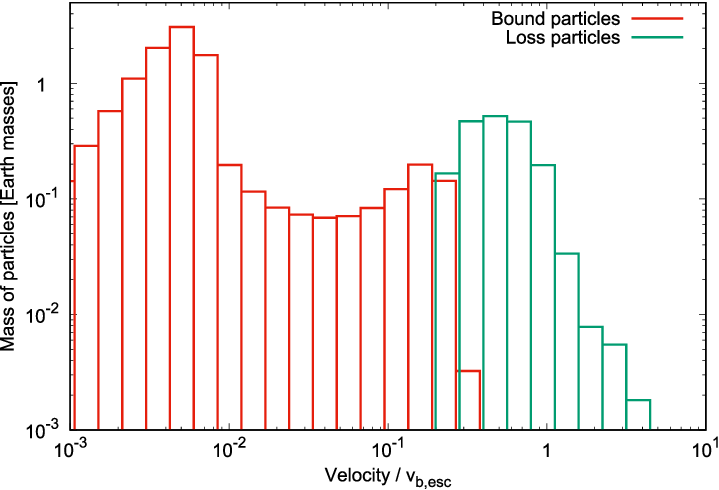}
\caption{
The histogram of the sum of particles masses in a specific velocity range for Run~31. The velocity is normalized by Eq.~\ref{vesc-rad}. Red histograms are bound particles. Green histograms are escaping particles. 
\label{hist_esc_vbesc}}
\end{center}
\end{figure}

Here we show the velocity distribution of particles. We show the Run~31 result.
Figure~\ref{vel_rad} shows the velocity distribution of particles after the collision. The escape velocity is considered to be the escape velocity at the location of the particles;
\begin{equation}
V_\mathrm{esc} = \sqrt{\frac{2GM_B}{D}} \label{vesc-rad}
\end{equation}
where $D$ is the distance from the center of the gravitationally bound body and $M_B$ is the mass of the gravitationally bound body.
We use $M_B$ from the simulation result of Run~31.
The atmospheric particle distribution beyond the escape velocity extends to about 10 times the original planetary radius.
The faster the particles are escaped by the impact, the further they will have travelled by this point.
The velocity distribution of the escaping particles revealed that the velocity of escaping particles are several times higher than $V_\mathrm{esc}$.
Since $V_\mathrm{esc}$ is inversely proportional to distance, the velocity of the escaping particles is faster than the escape velocity.
Thus, $V_\mathrm{esc}$ may underestimate the total kinetic energy of the escaping particles if all of the escaping particles is assumed to be the escape velocity at the location of the particles.
Next, we examine that the particle velocity distribution is normalized by the escape velocity at the surface of a gravitationally bound body that is expressed by
\begin{equation}
v_\mathrm{B,esc} = \sqrt{\frac{2GM_B}{R(M_B,Y_B)}} \label{vbesc-def}
\end{equation}
where $R$ is the effective radius for the planet whose mass and atmospheric mass fractions are $M_B$ and $Y_B$ respectively. $R$ is determined by Eq.~\ref{Rb-def} in this study.
Figure~\ref{hist_esc_vbesc} shows the particle velocity distribution using Eq.~\ref{vbesc-def}. The results show that most escaping particles are ejected at a velocity of about $v_\mathrm{B,esc}$ or less. Thus, we estimate the total kinetic energy of the escaping particles by representing their velocity in terms of $v_\mathrm{B,esc}$. In section~\ref{sec42} we consider the kinetic energy of escaping particles by Eq.~\ref{vbesc-def}.

\bibliography{sample631}{}
\bibliographystyle{aasjournal}

\end{document}